\documentclass[aps,prb,twocolumn,superscriptaddress,longbibliography,epsfig]{revtex4-2}

\usepackage{amsmath,amssymb,amsfonts}
\usepackage{hyperref}
\usepackage{booktabs}
\usepackage{makecell}
\usepackage{diagbox}
\usepackage{graphicx}
\usepackage{array}
\usepackage{bm}
\usepackage{xcolor}
\usepackage{physics}
\usepackage{braket}

\newcommand{\Ztwo}{\mathbb{Z}_2}
\newcommand{\Zthree}{\mathbb{Z}_3}
\newcommand{\Zfour}{\mathbb{Z}_4}

\newcommand{\ie}{\textit{i.e.}}

\begin{document}

\title
{State-selective dynamical freezing in driven Rydberg atom chains}

\author{Madhumita Sarkar}
\thanks{These authors contributed equally to this work.}
\author{Sougato Bose}
\address{Department of Physics and Astronomy, University College London, Gower Street, London WC1E 6BT, United Kingdom}
\author{Roopayan Ghosh$^{\ast}$}
\email{roopayan@iitbbs.ac.in}
\address{School of Basic Sciences, Indian Institute of Technology Bhubaneswar, Bhubaneswar 752050, India}

\begin{abstract}
Dynamical freezing, the near-complete suppression of dynamics at special drive frequencies, offers a route to mitigating Floquet heating in periodically driven quantum systems. In interacting systems, however, the energy cost of each transition depends on the many-body configuration, so that the freezing condition itself becomes state dependent. We study this dependence in periodically driven Rydberg atom chains under single- and bi-frequency driving. Using Floquet perturbation theory, we show that each spin flip is characterised by a single interference amplitude set by its local interaction shift, and that a state freezes only where the amplitudes of all its available flips vanish simultaneously. This picture accounts quantitatively for the shift and proliferation of freezing frequencies with interatomic spacing and for the response of entire families of initial states, which are sorted by a single local motif. Freezing is strong only within a window of spacings bounded by interaction tails on one side and nearest-neighbour excitation facilitation on the other, outside of which the advantage of bi-frequency driving is lost. In the regime of large detuning amplitude, where freezing points are numerous and the perturbative description is controlled, this window leaves almost no room for the PXP limit.
\end{abstract}
\maketitle
\vspace{2pc}

\section{Introduction}
\label{sec:intro}

Periodic driving provides a versatile means of controlling quantum matter away from equilibrium, allowing effective Hamiltonians and non-equilibrium phases to be engineered beyond what is accessible in static systems~\cite{Bukov2015,Eckardt2017,Oka2019,Sen2021JPCM,Zhang2017DTC,Choi2017DTC,Mi2022DTC}. A fundamental obstacle, however, is Floquet heating: generic interacting systems continuously absorb energy from the drive and eventually approach a featureless infinite-temperature state~\cite{DAlessio2014,Lazarides2014,Ponte2015}. While high-frequency driving can suppress heating and give rise to long-lived pre-thermal regimes~\cite{Abanin2015,Mori2016,Kuwahara2016,Peng2021}, identifying mechanisms that inhibit energy absorption at moderate and low frequencies, where exotic non-equilibrium states with no equilibrium counterpart can be prepared is an important challenge.

Dynamical freezing, closely related to dynamical localization, is a phenomenon in which the dynamics of a periodically driven quantum system become strongly suppressed at special values of the driving parameters. It originates from coherent interference between different transition pathways and has been studied extensively in both single-particle and many-body systems~\cite{Dunlap1986,Grossmann1991,Lignier2007,Das2010,Bhattacharyya2012,Hegde2014}. In many-body systems, such suppression is often associated with the emergence of approximately conserved quantities that inhibit transitions between different configurations~\cite{Haldar2021, Guo2025}. Its manifestations have been explored in integrable and non-integrable spin chains, fermionic and quasiperiodic systems, Rydberg arrays, chaotic quantum dots and multi-step driving protocols~\cite{Kar2016,Ghosh2018,Mukherjee2020vacuum,Ghosh2020FPT,Banerjee2025flatband,Ghosh2025tworate,Lu2026}. By suppressing energy absorption, dynamical freezing can prolong non-equilibrium dynamics and mitigate Floquet heating.

While dynamical freezing can be exact in non-interacting and certain integrable systems, its stability in generic interacting systems is less straightforward. Interactions introduce additional many-body transition processes that can destabilise the interference conditions underlying freezing. In particular, the emergent conservation laws associated with freezing are generally valid only up to finite orders in perturbation theory, with higher-order processes, including many-body resonances and instanton-like tunnelling events, eventually leading to its breakdown~\cite{Mukherjee2026instanton,Haldar2024,Banerjee2026pairproduction}. Interactions can also modify the freezing conditions themselves through their dependence on the many-body configuration. In particular, the energy cost of a spin flip depends on the occupations of the surrounding sites, resulting in configuration-dependent effective detunings and, consequently, distinct interference conditions for different initial states. The resulting freezing response is therefore governed not only by the interaction strength but also by the spatial structure of the initial configuration. Multi-frequency driving introduces further complexity through additional harmonics and the associated multi-photon transition pathways~\cite{Else2020,Dumitrescu2022}. The interplay of these effects raises the question of how interactions and drive harmonics shape the initial-state dependence of dynamical freezing in many-body systems.

Programmable Rydberg atom arrays provide a natural setting to investigate these effects~\cite{Saffman2010,Browaeys2020}. Atoms trapped in optical tweezers and coupled to highly excited Rydberg states realise quantum spin models with tunable long-range van der Waals interactions and flexible geometries. These platforms have enabled the exploration of quantum many-body scars~\cite{Bernien2017,Turner2018,Bluvstein2021}, exotic quantum phases~\cite{Ebadi2021,Semeghini2021}, quantum critical dynamics~\cite{Keesling2019} and prethermal behaviour under periodic driving~\cite{Darbha2025}. In parallel, theoretical studies of Rydberg arrays and their constrained descriptions have uncovered the mechanisms behind weak ergodicity breaking and many-body scars~\cite{Turner2018PRB,Ho2019,Khemani2019,Lin2019,Choi2019,Serbyn2021}, the control of scarred and vacuum dynamics by periodic driving~\cite{Mukherjee2020scars,Mukherjee2020vacuum,Maskara2021,Hudomal2022}, and a rich variety of ordered phases and transitions in chains, ladders and two-dimensional arrays~\cite{Chepiga2021,Samajdar2020,Verresen2021,Sarkar2023ladder,Pal2025scar}. 

In the absence of driving, a Rydberg chain realises an Ising model in transverse and longitudinal fields with long-range interactions, reducing to the constrained PXP model in the strong-blockade limit~\cite{Fendley2004,Lesanovsky2012,Turner2018}. The independent modulation of the detuning and Rabi frequency, together with the tunability of the interatomic spacing, makes these systems particularly suitable for investigating the interplay between multi-frequency driving and interactions. Related phenomena can also be explored in other programmable quantum platforms, including trapped ions~\cite{Zhang2017DTC,Dumitrescu2022}, superconducting qubits~\cite{Mi2022DTC}, nitrogen-vacancy centres~\cite{Choi2017DTC}, ultracold atoms in optical lattices~\cite{Lignier2007,Eckardt2017} and nuclear spins~\cite{Hegde2014,Peng2021}. The availability of cloud-accessible analogue simulators further broadens the scope for investigating such driven dynamics~\cite{Wurtz2023}.
In our recent work~\cite{Sarkar2026harmonic}, we experimentally demonstrated dynamical freezing of the vacuum state in Rydberg arrays of up to $100$ atoms and showed that bi-frequency modulation of the detuning and Rabi frequency suppresses dominant interaction-induced heating channels, substantially broadening the freezing regime. 

Here, we investigate how dynamical freezing depends on the initial many-body configuration and how this dependence evolves with interaction strength. We study the return fidelity of product states under single- and bi-frequency driving, considering the vacuum, density-wave states of different periodicities ($\Ztwo$, $\Zthree$ and $\Zfour$), and complete families of configurations with fixed excitation density, as the interatomic spacing is tuned continuously from the regime of strong interaction tails to that in which the blockade breaks down. Using Floquet perturbation theory~\cite{Ghosh2020FPT,Sen2021JPCM,Sarkar2026harmonic}, we show that each spin flip is characterised by a single interference amplitude set by its local interaction shift, and that a state freezes only where the amplitudes of all its available flips are suppressed simultaneously. In the PXP model, where all shifts vanish, this reduces to the single-atom freezing condition, independent of the initial state. With realistic van der Waals interactions, the shifts move the freezing frequencies continuously with spacing, generate dense combs of narrow resonances, and sort entire families of initial states according to a single local motif. Bi-frequency driving improves freezing as long as the relevant shifts remain small compared with the detuning amplitude, and loses its advantage otherwise. We compare the analytical predictions throughout with exact numerical simulations.

The rest of the paper is organised as follows. In Sec.~\ref{sec:model} we introduce the model, driving protocols and observables. Sec.~\ref{sec:fpt} develops the Floquet perturbation theory. In Secs.~\ref{sec:single} and~\ref{sec:bifreq} we present our results for single- and bi-frequency driving of ordered initial states, respectively, and compare them with the analytical predictions. In Sec.~\ref{sec:families} we turn to complete families of initial states with fixed excitation density and follow the freezing landscape continuously as a function of the interatomic spacing. We discuss our results in Sec.~\ref{sec:discussion}, and technical details are relegated to the appendices.

\section{Model and drive protocols}
\label{sec:model}
 
\subsection{Rydberg chain}
\label{sec:model:ham}
 
We consider a one-dimensional chain of $L$ neutral atoms with lattice spacing $a$ and open boundary conditions. Each atom is treated as a two-level system consisting of the electronic ground state $\ket{g}$ and a highly excited Rydberg state $\ket{r}$, coupled by a laser with time-dependent Rabi frequency $\Omega(t)$ and detuning $\Delta(t)$~\cite{Saffman2010,Browaeys2020}. The dynamics is governed by
\begin{equation}
H(t) = -\Delta(t)\sum_{i=1}^{L} n_i + \frac{\Omega(t)}{2}\sum_{i=1}^{L}\sigma^x_i + \sum_{i<j} V_{ij}\, n_i n_j ,
\label{eq:H}
\end{equation}
where $n_i=\ket{r}_i\!\bra{r}$, $\sigma^x_i=\ket{r}_i\!\bra{g}+\ket{g}_i\!\bra{r}$, and
\begin{equation}
V_{ij} = \frac{C_6}{\left(a\,|i-j|\right)^6}
\label{eq:vdw}
\end{equation}
is the van der Waals interaction between two Rydberg atoms. We take $C_6 = 5.42\times10^{6},\mu\mathrm{m}^6\mathrm{rad}/\mu\mathrm{s}$~\cite{Wurtz2023,Sarkar2026harmonic}. Writing $n_i=(1+\sigma^z_i)/2$, eq.~\eqref{eq:H} becomes, up to a constant,
\begin{equation}
H(t) = \frac{\Omega(t)}{2}\sum_i \sigma^x_i + \sum_i h_i(t)\,\sigma^z_i + \sum_{i<j} J_{ij}\,\sigma^z_i\sigma^z_j ,
\label{eq:Ising}
\end{equation}
with $J_{ij}=V_{ij}/4$ and $h_i(t) = -\Delta(t)/2 + \sum_{j\neq i}V_{ij}/4$. The Rydberg chain is therefore an Ising chain with long-range, rapidly decaying couplings in a transverse field $\Omega(t)/2$ and a site-dependent longitudinal field $h_i(t)$, both of which can be modulated in time.
 
The interaction Eq.~\eqref{eq:vdw} introduces a hierarchy of energy scales whose ordering depends sensitively on the spacing. We consider $a = 4.6$--$8\,\mu$m, over which the nearest-neighbour interaction $V_{i,i+1}$ decreases from about $570$ to $21\,\mathrm{rad}/\mu$s. For $a\leq6\,\mu$m it exceeds every other scale in the problem by at least an order of magnitude, and in particular remains far above the detuning amplitude $\Delta_0$ introduced below, so that simultaneous excitation of neighbouring atoms is energetically forbidden throughout the drive. This is the Rydberg blockade~\cite{Saffman2010,Browaeys2020}. In this regime we impose it exactly by restricting the dynamics to the constrained Hilbert space in which no two neighbouring atoms are excited, whose dimension grows as the Fibonacci number $F_{L+2}$~\cite{Fendley2004,Turner2018}; we have checked against simulations in the full Hilbert space that the constraint changes the fidelity by less than $5\times10^{-3}$. The remaining couplings, the interaction tails, are much weaker but by no means negligible: for $a=4.6$--$6\,\mu$m the next-nearest-neighbour coupling $V_{i,i+2}\simeq 1.8$--$9\,\mathrm{rad}/\mu$s is comparable to the drive frequency $\omega$ and to a sizeable fraction of $\Delta_0$, while $V_{i,i+3}\lesssim 0.8\,\mathrm{rad}/\mu$s is small but still of the order of the Rabi frequency. Changing $a$ by a fraction of a micron therefore tunes the tails through the scales of the drive. At $a=8\,\mu$m the tails beyond the nearest neighbour become negligible, $V_{i,i+2}\approx0.3\,\mathrm{rad}/\mu$s, but the hierarchy breaks down from the other side: $V_{i,i+1}$ is now comparable to $\Delta_0$, so that near the turning points of the drive, where $\Delta(t)\approx\Delta_0$, exciting an atom next to an existing excitation becomes nearly resonant. This is the facilitation, or anti-blockade, condition~\cite{Browaeys2020}. The nearest-neighbour interaction then acts as an ordinary, finite Ising coupling rather than a constraint, and for $a=8\,\mu$m we work in the full $2^L$-dimensional Hilbert space, keeping all interactions including $V_{i,i+1}$. The spacing thus interpolates between a strongly blockaded chain with appreciable tails and a nearest-neighbour Ising chain without blockade, and, as we show below, neither end of this range realises the PXP limit for the drive parameters used here.
 
\subsection{PXP limit}
\label{sec:model:pxp}
 
Retaining only the blockade and discarding all interaction tails, $V_{i,i+1}\to\infty$ and $V_{ij}\to0$ for $|i-j|\geq2$, yields the PXP model~\cite{Fendley2004,Lesanovsky2012,Turner2018}
\begin{equation}
H_{\rm PXP}(t) = -\Delta(t)\sum_i n_i + \frac{\Omega(t)}{2}\sum_i P_{i-1}\,\sigma^x_i\,P_{i+1},
\label{eq:PXP}
\end{equation}
where $P_i = 1-n_i$ projects onto the ground state of atom $i$. The PXP model has been remarkably successful in describing Rydberg chains, most notably in accounting for quantum many-body scars~\cite{Bernien2017,Turner2018,Bluvstein2021}, and its periodically driven version shows dynamical freezing of the vacuum and related states~\cite{Mukherjee2020vacuum,Mukherjee2020scars}.
 
For the questions we ask here, however, the PXP limit is special in an important way. Within the constrained space, the energy cost of flipping atom $i$ is $\mp\Delta(t)$ irrespective of the state of every other atom: once a flip is allowed by the constraint, it costs exactly the same energy as it would for an isolated atom. Every allowed flip therefore accumulates the same dynamical phase over a drive period, and the interference condition that determines dynamical freezing is identical for all of them. The constraint restricts \emph{which} flips are possible, but not \emph{how} their amplitudes interfere. In the PXP model, the freezing frequencies are consequently fixed by the single-atom drive alone, in the spirit of coherent destruction of tunnelling~\cite{Grossmann1991,Dunlap1986}, and are insensitive to the initial state, up to corrections that appear only beyond leading order in the drive (section~\ref{sec:fpt}).
 
The interaction tails remove this degeneracy. In the full model~\eqref{eq:H}, flipping atom $i$ costs in addition
\begin{equation}
\alpha_i = \sum_{|j-i|\geq 2} V_{ij}\, n_j ,
\label{eq:alpha}
\end{equation}
the interaction energy with the Rydberg excitations already present outside the blockade radius. The phase accumulated by a flip thus depends on the many-body configuration of the surrounding atoms, so different processes interfere differently, and the freezing condition itself becomes configuration-dependent. This is the origin of the interaction- and initial-state dependence studied in this paper, and it is absent by construction in the PXP model, which behaves as a model with an infinite nearest-neighbour interaction and nothing else. For the vacuum, $\alpha_i=0$ for every first flip, so the tails enter only through virtual processes involving two or more excitations; for states with a finite density of excitations, they enter already at the level of single flips (see section~\ref{sec:fpt}).
 
\subsection{Drive protocols}
\label{sec:model:drive}
 
In all cases the detuning is modulated sinusoidally,
\begin{equation}
\Delta(t) = \Delta_0\cos(\omega t),
\label{eq:detuning}
\end{equation}
so that over each period the chain is swept twice through the regime of small detuning, where flips are resonant. We compare two protocols for the Rabi frequency. In the \emph{single-frequency} protocol it is constant, $\Omega(t)=\bar\Omega$. In the \emph{bi-frequency} protocol it is modulated at the second harmonic~\cite{Sarkar2026harmonic},
\begin{equation}
\Omega(t) = \bar\Omega\left[1+\cos(2\omega t)\right],
\label{eq:bifreq}
\end{equation}
so that the Rabi coupling is maximal at $t=0, T/2$, when the detuning is largest, and vanishes at $t=T/4, 3T/4$, when the detuning passes through zero. Both protocols have the same time-averaged Rabi frequency $\bar\Omega$, so that any difference between them reflects the harmonic content of the drive rather than its overall strength. We use $\Delta_0 = 20\,\mathrm{rad}/\mu$s, $\bar\Omega = 1\,\mathrm{rad}/\mu$s and drive frequencies $\omega\in[1.5, 4.5]\,\mathrm{rad}/\mu$s, comparable to recent experiments on the QuEra Aquila device~\cite{Sarkar2026harmonic,Wurtz2023}. In the notation of Ref.~\cite{Sarkar2026harmonic}, the bi-frequency protocol corresponds to $\Omega(t)=\frac{\Omega_0}{2}[1+\cos(2\omega t)]$ with $\Omega_0=2\,\mathrm{rad}/\mu$s.
 
\subsection{Initial states and observables}
\label{sec:model:obs}
 
We prepare the chain in a product state $\ket{\psi_0}$ in the occupation basis and record the return fidelity after one drive period $T=2\pi/\omega$,
\begin{equation}
F(T) = \left|\bra{\psi_0}U(T)\ket{\psi_0}\right|^2 ,
\label{eq:fid}
\end{equation}
where $U(T)$ is the Floquet operator over one period. Dynamical freezing corresponds to $F(T)\to1$, so we plot $\ln[1-F(T)]$, in which freezing appears as a sharp minimum. We consider the vacuum $\ket{0}=\ket{gg\cdots g}$ and the density-wave states $\ket{\mathbb{Z}_k}$, in which every $k$th atom is excited, for $k=2,3,4$. These form a sequence with excitation density $1/k$ in which the distance between neighbouring excitations, and hence the dominant interaction tail they experience, grows with $k$. For $\mathbb{Z}_3$ we restrict to system sizes that are multiples of three, so that the pattern fits the chain exactly. We take the pattern with the first atom excited, $\ket{100100\cdots100}$; its last atom is then not blockaded and lies at a distance $2a$ from an excitation, so that it carries the interaction shift of a $\mathbb{Z}_2$ flip rather than that of the bulk (section~\ref{sec:fpt:first}). This single edge flip contributes noticeably to the infidelity close to the freezing points, which is a further manifestation of the environment dependence discussed below; in the maps of section~\ref{sec:families:contour} we instead use the pattern $\ket{010010\cdots010}$, in which both end atoms are blockaded. To go beyond these ordered states, we also consider all product states with a fixed number $n$ of excitations, and study both the fidelity of individual states and its average over each family.
 
We obtain $U(T)$ by numerically exact time evolution in the constrained Hilbert space, working in the interaction picture with respect to the diagonal part of~\eqref{eq:H} and integrating the remaining equations with a fourth-order Runge--Kutta scheme. We reach system sizes up to $L=18$ for $\ket{0}$, $\ket{\mathbb{Z}_2}$ and $\ket{\mathbb{Z}_4}$, and $L=21$ for $\ket{\mathbb{Z}_3}$, in the constrained space; for $a=8\,\mu$m, where the full Hilbert space is required, we reach $L=16$. The time step is chosen such that $\ln(1-F)$ is converged to better than $10^{-5}$ (see appendix~\ref{app:numerics} for more details).

\section{Floquet perturbation theory}
\label{sec:fpt}

To understand which features of the freezing landscape are set by the drive and which by the interactions, we construct the Floquet Hamiltonian $H_F$, perturbatively in the Rabi frequency~\cite{Ghosh2020FPT,Sen2021JPCM,Sarkar2026harmonic}. We treat the detuning and the interactions exactly, which resums them to all orders in $\Delta_0/\omega$ and $V_{ij}/\omega$, and expand only in the off-diagonal drive. For single-frequency driving the detuning sweeps through resonance twice per period with the drive switched on, and the expansion parameter is the Landau--Zener amplitude per passage, $\bar\Omega/\sqrt{\Delta_0\omega}$, set by the time $(\Delta_0\omega)^{-1/2}$ spent near each crossing. For bi-frequency driving the Rabi coupling vanishes at the crossings, the resonant passages are strongly suppressed, and the expansion is controlled instead by the off-resonant admixture at the turning points of the sweep, of order $\bar\Omega/\Delta_0$. Both parameters are small throughout our parameter range, even where $\bar\Omega/\omega$ is not (appendix~\ref{app:fpt}). They grow when an interaction shift approaches $\Delta_0$, where the two crossings of a flip merge at a turning point of the sweep.

\subsection{Interaction picture}
\label{sec:fpt:ip}

We write $H(t)=H_0(t)+H_1(t)$, with the diagonal part
\begin{equation}
H_0(t) = -\Delta_0\cos(\omega t)\sum_i n_i + \sum_{i<j}V_{ij}\,n_in_j
\end{equation}
and the drive $H_1(t)=\frac{\bar\Omega}{2}f(t)\sum_i\tilde\sigma^x_i$. Here $f(t)=1$ for the single-frequency and $f(t)=1+\cos(2\omega t)$ for the bi-frequency protocol, and $\tilde\sigma^{x}_i$ is $\sigma^x_i$ projected onto the constrained space (no projection is needed in the full Hilbert space). Since $H_0(t)$ commutes with itself at all times, its evolution operator is diagonal in the occupation basis, $U_0(t)=\exp[\,i(\Delta_0/\omega)\sin(\omega t)\sum_in_i-it\sum_{i<j}V_{ij}n_in_j]$. Because $\sin(\omega T)=0$, the detuning drops out after a full period and the zeroth-order Floquet Hamiltonian is the static interaction energy, $H_F^{(0)}=\sum_{i<j}V_{ij}n_in_j$, which is diagonal and does not change the occupations.

In the interaction picture, $H_I(t)=U_0^\dagger(t)H_1(t)U_0(t)$, each spin flip acquires a phase set by the energy it costs. Raising atom $i$ costs $-\Delta(t)+\alpha_i$, where $\alpha_i$ in the blockaded picture is defined in equation~\eqref{eq:alpha}, and while working in the full space it includes the terms where $j=i\pm 1$. It
is the interaction energy with the excitations already present (in the constrained space the nearest neighbours of a flippable atom are necessarily empty, so only the tails contribute). Since $\alpha_i$ does not involve $n_i$, it is the same before and after the flip, and
\begin{align}
H_I(t)=\frac{\bar\Omega}{2}f(t)\sum_i\left[e^{i\phi_i(t)}\tilde\sigma^+_i+{\rm h.c.}\right],\nonumber \\
\phi_i(t)=\alpha_it-x\sin(\omega t),\qquad x=\frac{\Delta_0}{\omega}.
\label{eq:HI}
\end{align}
The operator $\alpha_i$ is diagonal and acts on whatever configuration the flip is applied to. All the many-body physics of the problem enters through it.

The one-period evolution operator factorises as $U(T)=U_0(T)\,U_I(T)$, where $U_0(T)=e^{-iH_F^{(0)}T}$ is diagonal and $U_I(T)=\mathcal{T}\exp[-i\int_0^TH_I(t)\,dt]$ is expanded in the drive,
\begin{align}
U_I(T)&=1+U_1+U_2+\dots,\qquad U_1=-i\int_0^T\!dt\,H_I(t),\nonumber\\
U_2&=-\int_0^T\!dt_1\int_0^{t_1}\!dt_2\,H_I(t_1)H_I(t_2).
\label{eq:Dyson}
\end{align}
Writing $U_I(T)=\exp[-i(H_F^{(1)}+H_F^{(2)}+\dots)T]$ and matching order by order gives
\begin{equation}
H_F^{(1)}=\frac{i}{T}\,U_1,\qquad H_F^{(2)}=\frac{i}{T}\left(U_2-\frac{1}{2}U_1^2\right).
\label{eq:HForders}
\end{equation}
Since $U_0(T)$ is diagonal in the occupation basis, it contributes only a configuration-dependent phase and drops out of the return probability of any product state and of all diagonal observables, which are therefore determined by $H_F^{(1)}+H_F^{(2)}$ alone.

\subsection{First order}
\label{sec:fpt:first}

From equations~\eqref{eq:Dyson} and ~\eqref{eq:HForders}, $H_F^{(1)}=T^{-1}\int_0^TH_I(t)\,dt$. Using the Jacobi--Anger expansion, the drive envelope and the detuning phase combine into a single harmonic series,
\begin{align}
&f(t)\,e^{-ix\sin(\omega t)}=\sum_{n=-\infty}^{\infty}B_n(x)\,e^{-in\omega t},\nonumber \\
&B_n(x)=\left\{\begin{array}{ll}J_n(x) & \mbox{single frequency},\\[2pt] J_n(x)+\frac12\left[J_{n-2}(x)+J_{n+2}(x)\right] & \mbox{bi-frequency},\end{array}\right.
\label{eq:Bn}
\end{align}
and the time integral can be done in closed form. We find
\begin{align}
&H_F^{(1)}=\frac{\bar\Omega}{2}\sum_i\left[\mathcal{I}(\alpha_i)\,\tilde\sigma^+_i+{\rm h.c.}\right],\nonumber \\
&\mathcal{I}(\alpha)=e^{i\pi\alpha/\omega}\,\frac{\sin(\pi\alpha/\omega)}{\pi}\sum_{n=-\infty}^{\infty}\frac{B_n(x)}{\alpha/\omega-n}.
\label{eq:I}
\end{align}
The amplitude $\mathcal{I}(\alpha)$ is the net result of the interference, over one period, of the two passages of the flip energy $-\Delta(t)+\alpha$ through zero, and it is the central object of this section. Its consequence for the fidelity is most transparent for a product initial state $\ket{c}$. As noted above, only $H_F^{(1)}$ appears in leading order as it is the lowest order term that flips atoms, $\bra{c}H_F^{(1)}\ket{c}=0$. Expanding $e^{-iH_F^{(1)}T}$ to second order then gives $1-F(T)\simeq T^2\bra{c}\big(H_F^{(1)}\big)^2\ket{c}$, \ie,
\begin{equation}
1-F(T)\simeq\left(\frac{\pi\bar\Omega}{\omega}\right)^2\sum_{i\in\mathcal{A}(c)}\left|\mathcal{I}\big(\alpha_i(c)\big)\right|^2,
\label{eq:F1}
\end{equation}
where $\mathcal{A}(c)$ is the set of atoms that can be flipped in $\ket{c}$ and $\alpha_i(c)$ is evaluated in that configuration. Equation~\eqref{eq:F1} holds as long as $1-F\ll1$, which is precisely the regime close to the freezing points that we are interested in. It says that a state freezes only if \emph{every} flip available to it is suppressed, \ie, if $\mathcal{I}(\alpha_i(c))\approx0$ for all $i\in\mathcal{A}(c)$ at the same drive frequency.

\paragraph{Vacuum and PXP limit.} When $\alpha=0$, only the $n=0$ term in equation~\eqref{eq:I} survives and $\mathcal{I}(0)=B_0(x)$. This applies to every flip out of the vacuum, and to every flip in the PXP model, where the tails are absent. The freezing condition is then the single-particle one: $J_0(\Delta_0/\omega)=0$ for single-frequency driving, and $J_0+J_2=2J_1(x)/x=0$ for bi-frequency driving~\cite{Sarkar2026harmonic}. These conditions fix the positions of the minima seen for the vacuum and in the PXP column of figures~\ref{fig:single} and~\ref{fig:bi}. Because the condition is the same for every flip, it is also independent of the initial state, which is why all four states freeze at the same frequencies in the PXP limit. The bi-frequency drive also changes the amplitude away from the zeros: $2J_1(x)/x$ decays as $x^{-3/2}$ rather than $x^{-1/2}$, so the vacuum is suppressed over the whole frequency window and not only at isolated points~\cite{Sarkar2026harmonic}.

\paragraph{Finite interaction shifts.} For $\alpha\neq0$, equation~\eqref{eq:I} mixes all harmonics, weighted by a kernel peaked at $n\approx\alpha/\omega$ with a width of order one. Two consequences follow. First, since $B_n(x)$ is exponentially small for $|n|>x$, the amplitude is appreciable only for $|\alpha|\lesssim\Delta_0$. A flip whose interaction shift exceeds the detuning amplitude never comes into resonance during the sweep and is dynamically blocked; one with $|\alpha|<\Delta_0$ is crossed twice per period and its interference can be either constructive or destructive. Second, the zeros of $\mathcal{I}(\alpha)$, and hence the freezing frequencies of the corresponding flip, move continuously with $\alpha/\omega$. Whenever $\alpha/\omega$ passes through an integer $m$, the flip is resonant with the $m$th harmonic and $\mathcal{I}=\pm B_m(x)$, whose zeros differ from those of $B_0$. For $\alpha/\omega\gg1$ the amplitude therefore oscillates rapidly with $\omega$, producing a dense comb of narrow features.

\paragraph{Environments of the initial states.} Equation~\eqref{eq:F1} reduces the freezing of a product state to a small number of local environments, each characterised by its shift $\alpha$. In the bulk of the constrained chain:
\begin{itemize}
\item the vacuum has a single class of flips, all with $\alpha=0$;
\item in $\ket{\mathbb{Z}_2}$ and $\ket{\mathbb{Z}_3}$ every ground-state atom is blockaded, and the only flips are de-excitations, with $\alpha\simeq2V(2a)$ and $\alpha\simeq2V(3a)$ respectively, where $V(r)=C_6/r^6$;
\item $\ket{\mathbb{Z}_4}$ has \emph{two} classes: de-excitations with $\alpha\simeq2V(4a)\approx0$, and excitations of the atom midway between two excitations, with $\alpha\simeq2V(2a)$.
\end{itemize}
Table~\ref{tab:alpha} lists these shifts, and three features of the numerics follow immediately. (i) For $\ket{\mathbb{Z}_3}$, $\alpha/\omega$ stays below unity at all spacings, so this state remains close to the vacuum and PXP behaviour, with smoothly shifted minima. (ii) For $\ket{\mathbb{Z}_2}$ at the smallest spacing, $\alpha/\omega$ sweeps through many integers across the frequency window, producing a dense comb of minima; at larger spacing, where $\alpha/\omega$ is of order one, the minima are shifted but broad. (iii) $\ket{\mathbb{Z}_4}$ must suppress both its classes at once. Its $\alpha\approx0$ class favours the vacuum freezing frequencies, while the $\alpha\simeq2V(2a)$ class favours those of $\ket{\mathbb{Z}_2}$, so it cannot freeze as well as either and inherits the comb of $\ket{\mathbb{Z}_2}$ at small spacing. Because $\alpha\propto a^{-6}$, a change of spacing by a fraction of a micron sweeps $\alpha/\omega$ through several integers, moving each freezing point between resonant and antiresonant values. This is the origin of the non-monotonic dependence on $a$. At the open edges, an atom has fewer neighbours and hence a different $\alpha$, which explains the commensurability effects for $\mathbb{Z}_3$ and $\mathbb{Z}_4$ (section~\ref{sec:model:obs}).

\begin{table*}
\centering
\renewcommand{\arraystretch}{1.25}
\begin{tabular}{|l|c|c|c|c|c|}
\hline
Flip class & $\alpha$ & $a=4.6\,\mu$m & $a=5.4\,\mu$m & $a=6\,\mu$m & $a=8\,\mu$m\\
\hline\hline
$\mathbb{Z}_2$, de-excitation & $2V(2a)$ & 18.2 [4.0--12] & 6.95 [1.5--4.6] & 3.69 [0.8--2.5] & 0.66 [0.15--0.44]\\
\hline
$\mathbb{Z}_3$, de-excitation & $2V(3a)$ & 1.60 [0.35--1.1] & 0.61 [0.14--0.41] & 0.32 [0.07--0.22] & 0.06 [0.01--0.04]\\
\hline
$\mathbb{Z}_4$, de-excitation & $2V(4a)$ & 0.28 [0.06--0.19] & 0.11 [0.02--0.07] & 0.06 [0.01--0.04] & 0.01 [$<$0.01]\\
\hline
$\mathbb{Z}_4$, mid-site excitation & $2V(2a)$ & 17.9 [4.0--12] & 6.84 [1.5--4.6] & 3.64 [0.8--2.4] & 0.65 [0.14--0.43]\\
\hline\hline
one excited neighbour & $V(a)$ & -- & -- & -- & 20.7 [4.6--14]\\
\hline
two excited neighbours & $2V(a)$ & -- & -- & -- & 41.4 [9.2--28]\\
\hline
\end{tabular}
\caption{\label{tab:alpha}Bulk interaction shifts $\alpha$ (in $\mathrm{rad}/\mu$s) of the available flips for the density-wave states, with the range of $\alpha/\omega$ over the drive window $\omega\in[1.5,4.5]\,\mathrm{rad}/\mu$s in brackets. The vacuum and all flips in the PXP model have $\alpha=0$. The last two rows apply only in the full Hilbert space (section~\ref{sec:fpt:cascade}).}
\end{table*}

\subsection{Higher orders: interaction-induced cascades}
\label{sec:fpt:cascade}

Equation~\eqref{eq:F1} treats every flip as if it took place in the unperturbed initial configuration. Beyond first order this is no longer true, because every flip changes the environment of all atoms within the interaction range: creating or removing an excitation at site $j$ shifts $\alpha_i\to\alpha_i\pm V_{ij}$. Inserting equation~\eqref{eq:HI} into equation~\eqref{eq:HForders}, at second order~\cite{Sarkar2026harmonic},
\begin{align}
H_F^{(2)}=&-\frac{i}{T}\left(\frac{\bar\Omega}{2}\right)^2\sum_{i,j}\sum_{s,s'=\pm}[K^{ss'}_{ij}\nonumber \\
&-\frac{T^2}{2}\,\mathcal{I}^{s}\!\left(\alpha_i+s'V_{ij}\right)\mathcal{I}^{s'}\!\left(\alpha_j\right)]\tilde\sigma^{s}_i\tilde\sigma^{s'}_j ,
\label{eq:HF2}
\end{align}
where $\mathcal{I}^{+}=\mathcal{I}$, $\mathcal{I}^{-}=\mathcal{I}^*$, the shifts are evaluated in the configuration before both flips, and the time-ordered kernel
\begin{align}
K^{ss'}_{ij}&=\sum_{m,n}B_m(sx)B_n(s'x)\nonumber \\ &\int_0^T\!dt_1\,e^{i[s(\alpha_i+s'V_{ij})-m\omega]t_1}\int_0^{t_1}\!dt_2\,e^{i[s'\alpha_j-n\omega]t_2}
\label{eq:K}
\end{align}
describes a flip of atom $j$ at time $t_2$ followed by a flip of atom $i$ at $t_1>t_2$. The second flip sees the environment modified by the first, $\alpha_i\to\alpha_i+s'V_{ij}$; the elementary integrals and their resonant limits are given in appendix~\ref{app:kernel}. We evaluate $H_F^{(1)}+H_F^{(2)}$ in the same Hilbert space as the exact numerics, truncating the harmonic sums at $|n|\leq24$, and verify that $e^{-iH_FT}$ agrees with the exact $U(T)$ up to corrections of third order in the expansion parameter introduced at the beginning of this section (appendix~\ref{app:fpt}).

Equation~\eqref{eq:HF2} makes precise how interactions spread through the dynamics. A flip that is frozen at first order because $\mathcal{I}(\alpha_j)=0$ still opens a virtual pathway in which a second atom $i$ is flipped with a shifted amplitude $\mathcal{I}(\alpha_i\pm V_{ij})$. This amplitude is generically not at a zero, and it becomes resonant when $V_{ij}$ is commensurate with a harmonic of the drive. For the vacuum this is the only way the interactions enter: all first flips have $\alpha=0$ and freeze exactly at the Bessel zeros, but the second flip, at distance $r$ from the first, sees $\alpha=V(r)$. The vacuum therefore retains the PXP freezing frequencies, with residual dynamics suppressed by a further factor of the square of the expansion parameter and growing with the tails, consistent with the weak but visible distance dependence of its minima. In states that already contain excitations, the same mechanism acts on top of a first-order landscape that is already split into several environment classes, and each flip diversifies the environments further. This is a cascade in the following sense: one flip can turn a blocked or frozen channel of a neighbour into an active one. Whether it does so depends on how the excitations are arranged, which is why families with the same number of excitations but different spacings respond so differently (section~\ref{sec:families}).

The full-space results at $a=8\,\mu$m make this explicit. There the nearest-neighbour coupling $V(a)$ is no longer a constraint but an ordinary shift, comparable to $\Delta_0$ and hence at the edge of the crossing window $|\alpha|<\Delta_0$ (table~\ref{tab:alpha}). A ground-state atom with exactly one excited neighbour has $\alpha\approx V(a)\approx\Delta_0$: its flip energy comes close to zero at the turning points of the sweep, where the bi-frequency Rabi coupling is largest, and it is facilitated. A ground-state atom between two excitations has $\alpha\approx2V(a)>\Delta_0$ and is dynamically blocked. Bulk $\ket{\mathbb{Z}_3}$ and $\ket{\mathbb{Z}_4}$ contain an extensive number of atoms of the first kind and lose their freezing. Bulk $\ket{\mathbb{Z}_2}$ contains only the second kind and is protected at first order: under bi-frequency driving, open chains of even length are destabilised by the single facilitated edge atom with one excited neighbour, while odd chains, with both ends excited, retain deeper minima; under single-frequency driving, with its weaker coupling at the turning points, this edge channel is only weakly active (appendix~\ref{app:blockade}). In the bulk the facilitated channel opens only through the cascade, when a de-excitation leaves a neighbour with a single excited partner. The vacuum has no excitations to facilitate from and is, to first order, unaffected.

\section{Single-frequency driving}
\label{sec:single}

\begin{figure*}
\centering
\includegraphics[width=\textwidth]{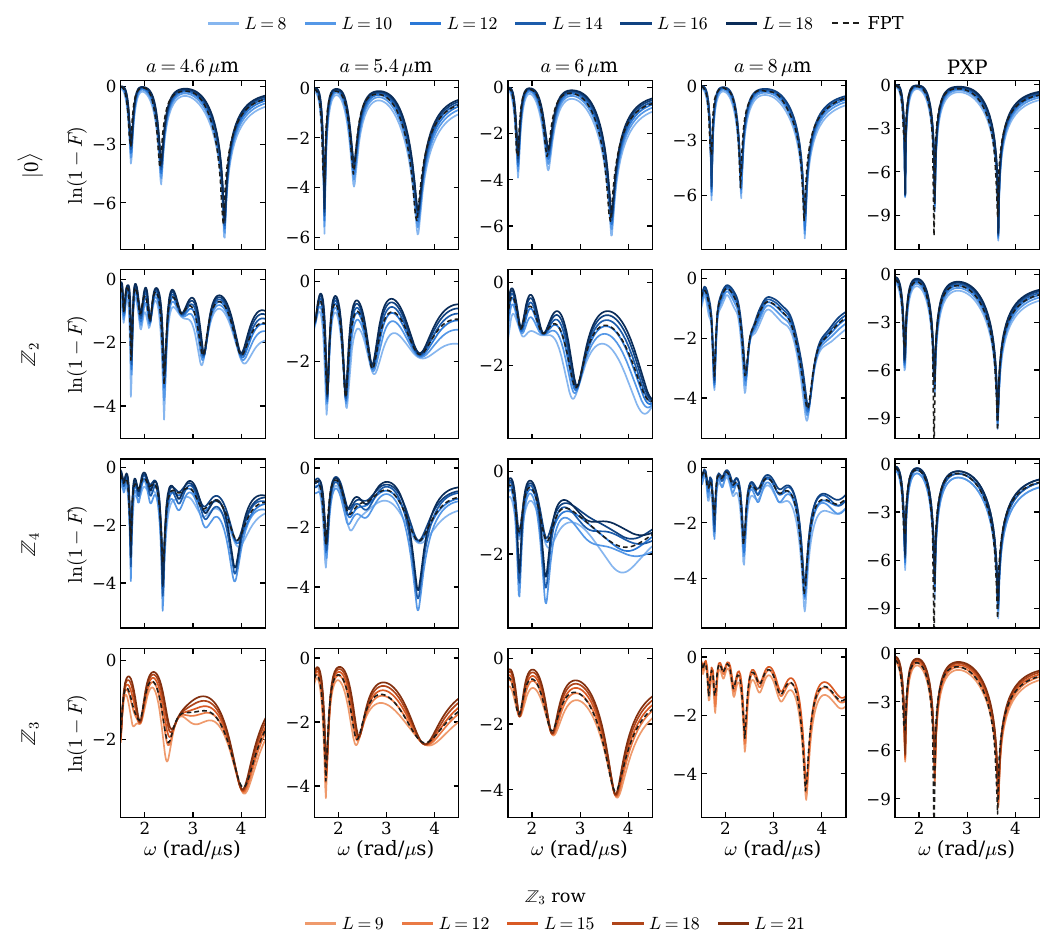}
\caption{\label{fig:single}Dynamical freezing of ordered initial states under single-frequency driving [$\Omega(t)=\bar\Omega$, equation~\eqref{eq:H} with~\eqref{eq:detuning}]. The logarithmic infidelity $\ln[1-F(T)]$ after one drive period [equation~\eqref{eq:fid}] is shown as a function of the drive frequency $\omega$, with freezing appearing as sharp minima. Rows: the vacuum $\ket{0}$, $\ket{\mathbb{Z}_2}$ and $\ket{\mathbb{Z}_4}$ (blue, $L=8$--$18$, darker for larger $L$), and $\ket{\mathbb{Z}_3}$ (orange, commensurate sizes $L=9$--$21$). Columns: lattice spacings $a=4.6$, $5.4$, $6$ and $8\,\mu$m, and the PXP limit [equation~\eqref{eq:PXP}]. All columns except $a=8\,\mu$m are computed in the blockade-constrained space; for $a=8\,\mu$m the full Hilbert space is used ($L\leq16$). Dashed lines: second-order Floquet perturbation theory for $L=12$.}
\end{figure*}

We begin with the single-frequency protocol, for which the Rabi frequency is constant and the only time dependence enters through the detuning. Figure~\ref{fig:single} shows the log infidelity after one period for the four ordered initial states, across the full range of spacings and in the PXP limit.

\paragraph{PXP limit.} In the PXP model all four states freeze at the same three frequencies, which coincide with the zeros of $J_0(\Delta_0/\omega)$. This is the behaviour anticipated in section~\ref{sec:fpt:first}: in the absence of interaction tails every available flip has $\alpha=0$, so the freezing condition is that of a single driven atom and carries no information about the initial configuration. The states differ only in the depth of the minima and in the background between them, reflecting the different number of flippable atoms. Between the minima the infidelity is of order one, so that under single-frequency driving freezing is confined to narrow windows around isolated frequencies.

\paragraph{Vacuum.} The vacuum retains the PXP freezing frequencies at all spacings, with minima whose depth varies with $a$ but whose positions do not move. Since every first flip out of the vacuum has $\alpha=0$, the tails enter only at second order, through processes in which a second excitation is created in the vicinity of the first (section~\ref{sec:fpt:cascade}). At $a=8\,\mu$m, where the blockade is no longer imposed, the vacuum is essentially indistinguishable from the PXP limit: with no excitations present, there is nothing to facilitate from.

\paragraph{Density-wave states.} States containing excitations depart from the PXP behaviour as soon as the interaction tails become appreciable, and they do so in a way that reflects the local environments identified in table~\ref{tab:alpha}. For $\ket{\mathbb{Z}_3}$, whose only flips carry a small shift $\alpha\simeq2V(3a)$, the minima remain close to their PXP positions but drift continuously to higher frequencies as the spacing decreases, while broadening and becoming shallower. This is the continuous motion of the zeros of $\mathcal{I}(\alpha)$ with $\alpha/\omega$ predicted by equation~\eqref{eq:I}. For $\ket{\mathbb{Z}_2}$, whose shift $\alpha\simeq2V(2a)$ is larger by roughly one order of magnitude, the effect is much stronger: already at $a=6\,\mu$m the high-frequency minimum is replaced by a single broad minimum at a different frequency, at $a=5.4\,\mu$m new minima appear between the PXP ones, and at $a=4.6\,\mu$m, where $\alpha/\omega$ passes through several integers across the frequency window, the low-frequency side develops a comb of narrow resonances. $\ket{\mathbb{Z}_4}$ illustrates the role of competing environments. Its de-excitations, with $\alpha\approx0$, are suppressed at the vacuum freezing frequencies, and its minima indeed remain pinned close to those positions at all spacings; its mid-site excitations, with $\alpha\simeq2V(2a)$, are not suppressed there, so the minima are uniformly shallower than for either the vacuum or $\ket{\mathbb{Z}_3}$. The ordering of the states is thus set by their environments rather than by their excitation density: $\ket{\mathbb{Z}_3}$, despite its higher density, freezes considerably better than $\ket{\mathbb{Z}_4}$.

\paragraph{Absence of the blockade.} At $a=8\,\mu$m the next-nearest-neighbour tails are negligible, and in the constrained space all three density-wave states should reproduce the PXP minima. However, in the full Hilbert space they differ from it. $\ket{\mathbb{Z}_2}$ keeps minima at the PXP positions, but shallower: its bulk ground-state atoms sit between two excitations and are dynamically blocked, and its infidelity is close to that of the constrained chain; the edge atom with a single excited neighbour, which is facilitated, contributes only weakly under single-frequency driving (appendix~\ref{app:blockade}). $\ket{\mathbb{Z}_3}$ and $\ket{\mathbb{Z}_4}$, which contain an extensive number of facilitated atoms, retain their principal minima (the minimas of the $PXP$ chain) but acquire additional weak resonances throughout the frequency window. Under single-frequency driving these facilitation processes remain comparatively mild, because the Rabi coupling at the turning points of the sweep, where the facilitation condition is approached, is no larger than at any other time. This will change for the bi-frequency drive.

\paragraph{System size and comparison with perturbation theory.} The positions of all minima are independent of $L$ within our resolution, and for $\ket{\mathbb{Z}_3}$ the commensurate sizes form a smooth family. The overall level of the curves rises with $L$, as expected from equation~\eqref{eq:F1}, according to which $1-F$ is a sum over the flippable atoms and hence grows with system size as long as it is small. For $\ket{\mathbb{Z}_4}$, the sizes $L=10$, $14$, $18$ are not commensurate with the period-four pattern, which modifies the configuration near one end of the chain; this produces a weak splitting between the two families of sizes, most visible at $a=6\,\mu$m, while leaving the positions of the minima unchanged. Second-order Floquet perturbation theory reproduces the exact curves in all panels, including the shifted and additional minima of the density-wave states and the full-space results at $a=8\,\mu$m, confirming that the structure of the freezing landscape is captured by the configuration-dependent amplitudes $\mathcal{I}(\alpha_i)$ and their second-order corrections.

\section{Bi-frequency driving}
\label{sec:bifreq}

\begin{figure*}
\centering
\includegraphics[width=\textwidth]{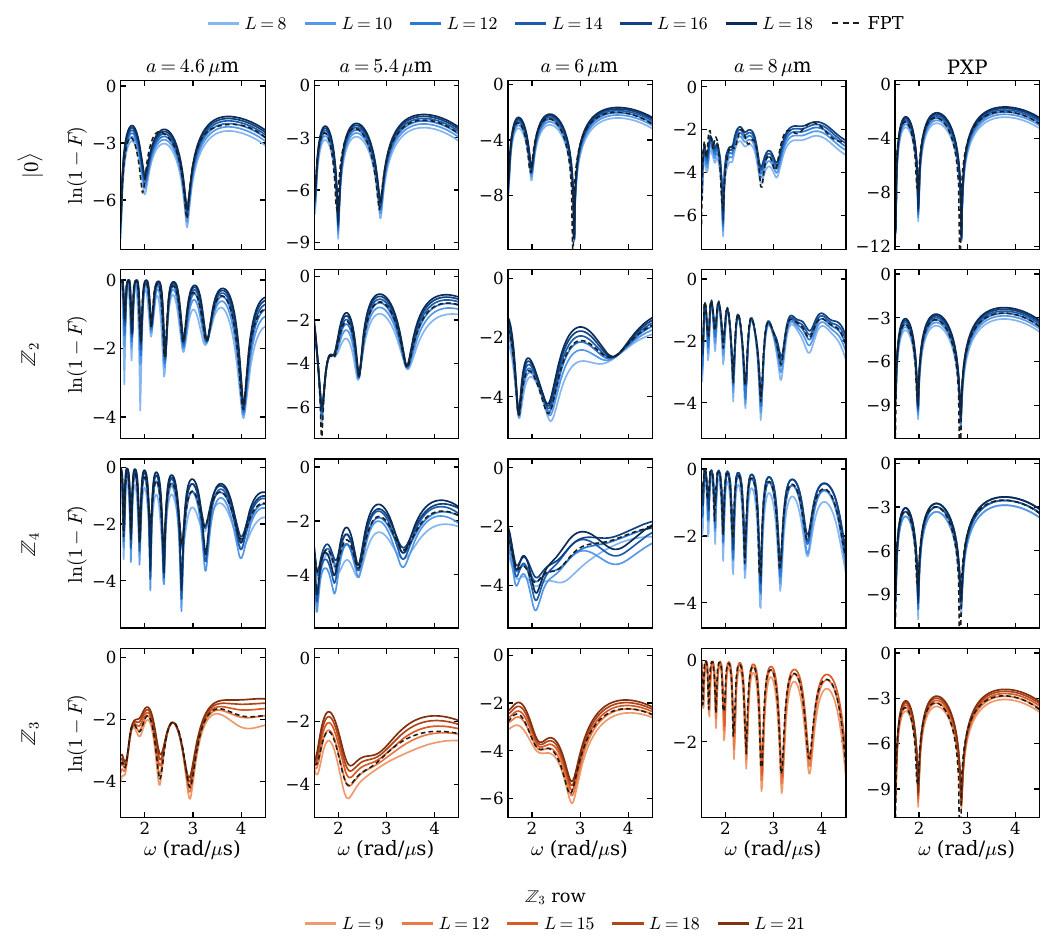}
\caption{\label{fig:bi}Dynamical freezing of ordered initial states under bi-frequency driving [$\Omega(t)=\bar\Omega[1+\cos(2\omega t)]$, equation~\eqref{eq:bifreq}], with the same time-averaged Rabi frequency as in figure~\ref{fig:single}. Layout, system sizes and line styles as in figure~\ref{fig:single}.}
\end{figure*}

\begin{figure*}
\centering
\includegraphics[width=\textwidth]{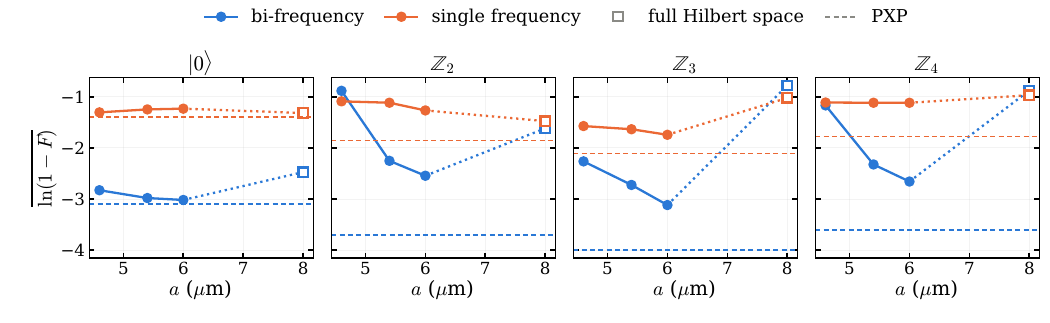}
\caption{\label{fig:summary}Infidelity averaged over the drive window $\omega\in[1.5,4.5]\,\mathrm{rad}/\mu$s, $\overline{\ln(1-F)}$, as a function of the lattice spacing for bi-frequency (blue) and single-frequency (orange) driving, for $L=16$ ($L=15$ for $\ket{\mathbb{Z}_3}$). Filled circles: blockade-constrained space; open squares: full Hilbert space at $a=8\,\mu$m; dotted lines connect the two. Dashed horizontal lines: PXP limit.}
\end{figure*}

We now add a second harmonic to the Rabi frequency, keeping its time average fixed, so that any difference from section~\ref{sec:single} reflects the harmonic content of the drive rather than its overall strength. The results are shown in figure~\ref{fig:bi}, and figure~\ref{fig:summary} compares the two protocols through the infidelity averaged over the drive window.

\paragraph{PXP limit and vacuum initial state.} In the PXP limit the minima move to the zeros of $2J_1(\Delta_0/\omega)/x$, again identically for all four states, and the infidelity is lowered across the entire frequency window rather than only near the minima. The vacuum behaves in the same way at every spacing in the constrained space, with its freezing frequencies fixed by the single-atom condition and the window-averaged infidelity close to its PXP value (figure~\ref{fig:summary}). For these states, bi-frequency driving converts the isolated freezing points of the single-frequency protocol into broad regions of strongly suppressed dynamics.

\paragraph{Why does the second harmonic help?} The origin of this improvement, and of its limits, becomes transparent once the bi-frequency coefficients of equation~\eqref{eq:Bn} are rewritten with the Bessel recurrence relations,
\begin{equation}
B_n(x)=\frac{2}{x}\left[\frac{n^2}{x}J_n(x)-J_n'(x)\right].
\label{eq:Bnbi}
\end{equation}
For $n=0$ this reduces to $2J_1(x)/x$. In the oscillatory regime $|n|<x$, the asymptotic forms of $J_n$ and $J_n'$ show that the envelope of $B_n$ is smaller than that of the single-frequency coefficient $J_n$ by a factor $(2/x^2)\sqrt{n^4+x^2-n^2}$. Since a flip with interaction shift $\alpha$ draws its amplitude from harmonics $n\approx\alpha/\omega$ [equation~\eqref{eq:I}], the second harmonic rescales its amplitude by
\begin{equation}
S(\alpha)\simeq\frac{2}{\Delta_0^2}\sqrt{\alpha^4+\omega^2\left(\Delta_0^2-\alpha^2\right)},\qquad|\alpha|<\Delta_0.
\label{eq:Salpha}
\end{equation}
For $|\alpha|\lesssim\sqrt{\omega\Delta_0}$ this gives $S\simeq2\omega/\Delta_0$, the same suppression as for the vacuum; for larger shifts it weakens as $2\alpha^2/\Delta_0^2$, reaches unity at $|\alpha|\simeq\Delta_0/\sqrt2$, and exceeds it as $|\alpha|$ approaches $\Delta_0$, where the bi-frequency drive enhances rather than suppresses the flip. The second harmonic is therefore effective for flips with $|\alpha|\lesssim\Delta_0/\sqrt2$, and becomes detrimental for flips close to the facilitation condition. This single criterion organises the behaviour of all the states we consider.

\paragraph{Density-wave states.} For $a=5.4$ and $6\,\mu$m, all flips of the density-wave states have $|\alpha|\ll\Delta_0$ (table~\ref{tab:alpha}), and bi-frequency driving improves their freezing substantially. The minima of $\ket{\mathbb{Z}_2}$, $\ket{\mathbb{Z}_3}$ and $\ket{\mathbb{Z}_4}$ remain shifted from the PXP positions, as dictated by their environments, but they deepen and merge into broad frequency ranges of low infidelity, most strikingly for $\ket{\mathbb{Z}_3}$ and $\ket{\mathbb{Z}_4}$ at $a=6\,\mu$m. At $a=4.6\,\mu$m the picture splits according to table~\ref{tab:alpha}. $\ket{\mathbb{Z}_3}$, whose shift is still small, retains a clear advantage from the second harmonic. $\ket{\mathbb{Z}_2}$ and the mid-site flips of $\ket{\mathbb{Z}_4}$, however, carry a shift $\alpha\simeq2V(2a)$ approaching $\Delta_0$; these flips are resonant with high harmonics, for which equation~\eqref{eq:Salpha} provides no suppression, and both states show the same dense comb of resonances as under single-frequency driving, with no gain in the averaged infidelity.

\paragraph{Absence of the blockade.} At $a=8\,\mu$m the same criterion applies to the nearest-neighbour coupling. The facilitated flips of $\ket{\mathbb{Z}_3}$ and $\ket{\mathbb{Z}_4}$ have $\alpha\approx V(a)\approx\Delta_0$ and receive no suppression from the second harmonic. Worse, the bi-frequency Rabi coupling is largest at the turning points of the sweep, precisely where the facilitation condition is approached. The infidelity of these states is consequently a dense comb of resonances, and bi-frequency driving performs no better, and on average worse, than single-frequency driving: the advantage of the second harmonic is reversed. $\ket{\mathbb{Z}_2}$, whose bulk atoms are dynamically blocked, is affected more weakly; here the facilitated edge atom produces a strong resonance, absent under single-frequency driving, while facilitated channels opened by de-excitations in the bulk account for the remainder (appendix~\ref{app:blockade}). Even the vacuum, which has no excitations to facilitate from, loses part of its bi-frequency advantage at this spacing. Since its first flips are unaffected, this arises at second order, through the creation of a nearest-neighbour pair~\cite{Banerjee2026pairproduction}, a two-photon process whose energy $-2\Delta(t)+V(a)$ vanishes during the sweep once $V(a)<2\Delta_0$. Close to the vacuum freezing frequencies we find that most of the population leaving the vacuum indeed ends up in configurations containing an adjacent pair for $a\gtrsim7\,\mu$m, whereas for smaller spacings it remains in blockade-allowed configurations. The onset of this channel coincides with the breakdown of the blockade marked in figure~\ref{fig:contour}.

\paragraph{Non-monotonic dependence on the spacing.} Figure~\ref{fig:summary} summarises these trends. Under single-frequency driving the window-averaged infidelity depends only weakly on the spacing, since freezing is in any case confined to narrow windows. Under bi-frequency driving it becomes strongly and non-monotonically dependent on $a$: every state freezes best at intermediate spacing, and freezing deteriorates both towards smaller spacing, where the next-nearest-neighbour tails reach the scale of $\Delta_0$, and towards larger spacing, where the nearest-neighbour coupling does. In both regimes the deterioration is strongest for states with excitations, because only they possess flips whose shift $\alpha$ can approach $\Delta_0$. The finite-size behaviour and the agreement with perturbation theory are as for single-frequency driving: the positions of the minima do not depend on $L$, and second-order Floquet perturbation theory reproduces the exact curves in all panels, including the full-space results at $a=8\,\mu$m.
\section{General initial states and continuous tuning of the spacing}
\label{sec:families}

\subsection{States with a fixed excitation density}
\label{sec:families:n}

\begin{figure*}
\centering
\includegraphics[width=\textwidth]{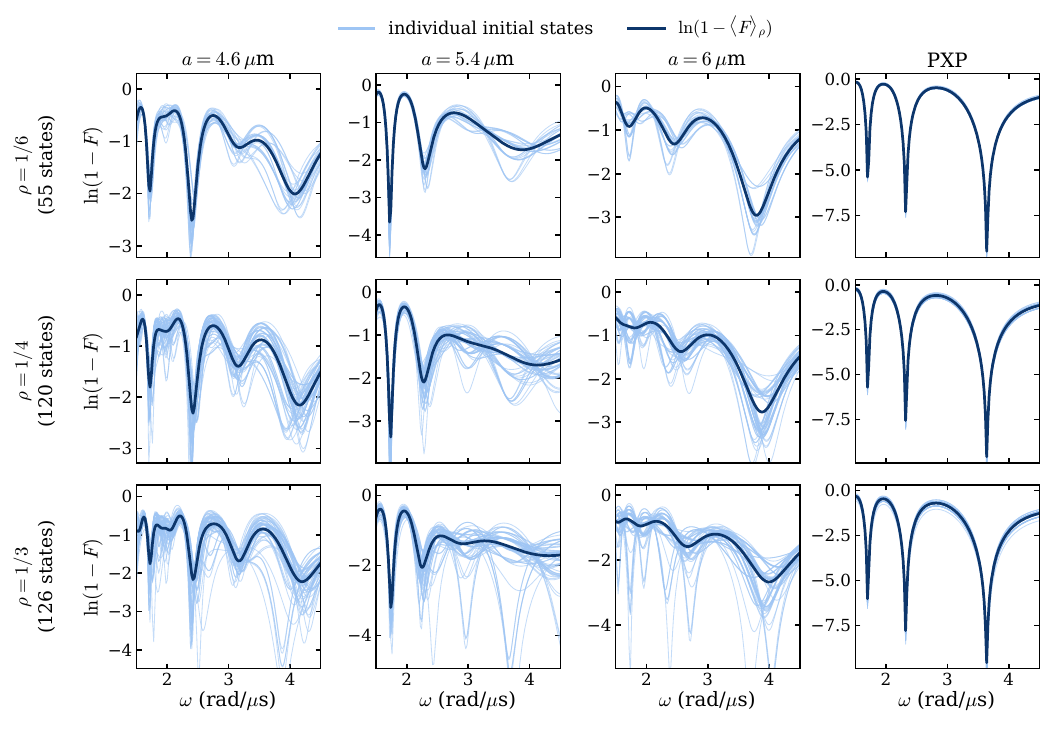}
\caption{\label{fig:fam_single}Single-frequency driving of all product states with a fixed number $n$ of excitations in a chain of $L=12$ atoms, corresponding to densities $\rho=n/L=1/6$, $1/4$ and $1/3$ (rows). Light lines show $\ln[1-F(T)]$ for the individual initial states and the dark line the logarithmic infidelity of the family average, $\ln(1-\langle F\rangle_\rho)$. Columns correspond to $a=4.6$, $5.4$ and $6\,\mu$m and to the PXP limit.}
\end{figure*}

\begin{figure*}
\centering
\includegraphics[width=\textwidth]{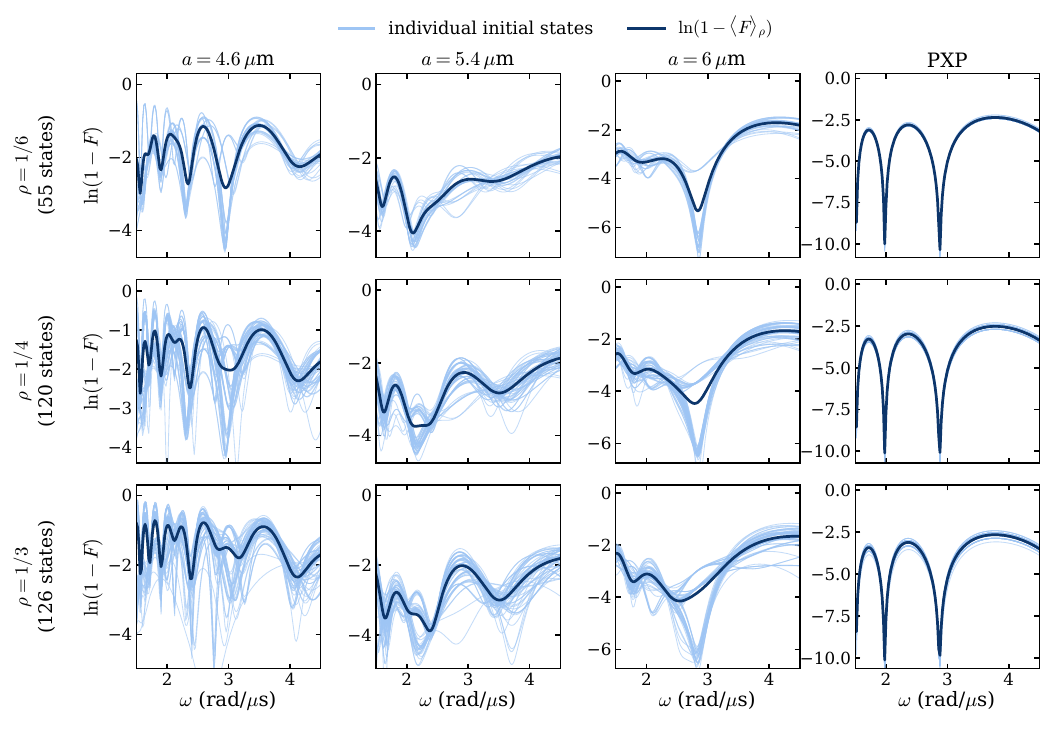}
\caption{\label{fig:fam_bi}Dynamical freezing of all product states with fixed excitation density under bi-frequency driving [$\Omega(t)=\bar\Omega[1+\cos(2\omega t)]$, equation~\eqref{eq:bifreq}], with the same time-averaged Rabi frequency as in figure~\ref{fig:fam_single}. Rows correspond to $n=2$, $3$ and $4$ excitations in a chain of $L=12$ atoms ($\rho=1/6$, $1/4$ and $1/3$), and columns to $a=4.6$, $5.4$ and $6\,\mu$m and to the PXP limit. Light lines: $\ln[1-F(T)]$ for the individual initial states; dark line: logarithmic infidelity of the family-averaged fidelity, $\ln(1-\langle F\rangle_\rho)$.}
\end{figure*}

\begin{figure}
\centering
\includegraphics[width=\columnwidth]{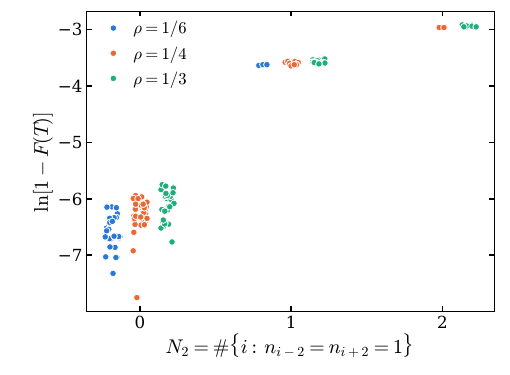}
\caption{\label{fig:motif}Infidelity of every state of the three families at the vacuum freezing frequency $\omega\simeq2.86\,\mathrm{rad}/\mu$s, for bi-frequency driving at $a=6\,\mu$m, as a function of the number $N_2$ of sites whose two next-nearest neighbours are both excited. Points are offset horizontally by density for clarity.}
\end{figure}

The density-wave states of sections~\ref{sec:single} and~\ref{sec:bifreq} are special in that all their excitations sit at the same distance from each other. To see how freezing depends on the arrangement of the excitations rather than on their number, we consider all product states with $n=2$, $3$ and $4$ excitations in a chain of $L=12$ atoms. This length is commensurate with all three densities $\rho=1/6$, $1/4$ and $1/3$, so that each family contains an evenly spaced member and all families share the same boundaries. Figures~\ref{fig:fam_single} and~\ref{fig:fam_bi} show the infidelity of every state in each family, together with the infidelity of the family-averaged fidelity $\langle F\rangle_\rho$.

\paragraph{PXP limit.} In the PXP model all states of all three families collapse onto essentially a single curve, with minima at the single-atom freezing frequencies. In the absence of interaction tails, neither the number nor the arrangement of the excitations enters the freezing condition, and the only remaining trace of the initial state is the number of flippable atoms, which sets the overall level of the curves.

\paragraph{A finite number of environments.} At finite spacing the states of a family separate, but the number of distinct environments they can provide does not grow with density. Because the interaction decays as $r^{-6}$, the shift of a flip in the constrained chain can be written as
\begin{equation}
\alpha_i=V(2a)\left[k_2+\left(\tfrac{2}{3}\right)^6k_3+\left(\tfrac{1}{2}\right)^6k_4+\dots\right],
\label{eq:alphak}
\end{equation}
where $k_r\in\{0,1,2\}$ is the number of excitations at distance $r$ from atom $i$. Since $(2/3)^6\approx0.09$, the shifts cluster into three bands, $\alpha\approx0$, $V(2a)$ and $2V(2a)$, labelled by $k_2$, whatever the density; the subleading terms only split these bands into finely spaced sub-levels, which matter when $V(3a)$ is no longer small compared with $\omega$, \ie, at the smallest spacing. Increasing the density does not create new bands but changes how many flips of a given configuration fall into each of them. Combining this with equation~\eqref{eq:F1}, the infidelity of a configuration $c$ is, to leading order, a weighted count,
\begin{equation}
1-F(T)\simeq\left(\frac{\pi\bar\Omega}{\omega}\right)^2\sum_{k=0}^{2}M_k(c)\,\left|\mathcal{I}\big(kV(2a)\big)\right|^2,
\label{eq:Fcount}
\end{equation}
where $M_k(c)$ is the number of flippable atoms with $k_2=k$. The freezing landscape of a whole family is therefore controlled by three amplitudes, and the individual states differ only in how they weight them.

\paragraph{Which states freeze.} Equation~\eqref{eq:Fcount} identifies which members of a family freeze at a given frequency. At the vacuum freezing frequencies $\mathcal{I}(0)$ vanishes, and for bi-frequency driving at $a=6\,\mu$m the $k_2=2$ band has an amplitude two orders of magnitude larger than the $k_2=1$ band, so that the infidelity is controlled by $M_2(c)$ alone. A site has $k_2=2$ precisely when both of its next-nearest neighbours are excited; such a site is always flippable, whether it is itself excited or not, so $M_2(c)=N_2(c)$ is simply the number of sites whose two next-nearest neighbours are both excited, $N_2(c)=\#\{i:\,n_{i-2}=n_{i+2}=1\}$. Figure~\ref{fig:motif} confirms that this single integer sorts all states of all three families: motif-free configurations retain the deep freezing of the vacuum, each occurrence of the motif adds an equal contribution to $1-F$, and states with the same $N_2$ freeze equally well irrespective of their density. The density enters only through the fraction of configurations that avoid the motif, which decreases as excitations are packed more closely. This is why the vacuum minimum survives in the family average at low density and is progressively washed out at higher density, while the spread among individual states grows.

\paragraph{Narrow resonances.} The same reasoning explains the minima that appear for states with excitations but are absent for the vacuum. A configuration whose flips all fall into a single band with $\alpha\neq0$ freezes wherever $\mathcal{I}(\alpha)$ vanishes, which is a different set of frequencies from the zeros of $B_0$. For $\ket{\mathbb{Z}_2}$, all bulk flips share the shift $\alpha\simeq2V(2a)$, and at $a=4.6\,\mu$m every minimum of the dense comb in figure~\ref{fig:bi} coincides with a zero of $\mathcal{I}(\alpha)$ for this shift. These zeros lie close to half-integer values of $\alpha/\omega$, where the interaction phase $\alpha T$ accumulated over one period is an odd multiple of $\pi$, so that successive minima are separated by $\Delta\omega\approx\omega^2/\alpha$. They are therefore densely spaced at low frequency and at small spacing, and narrow, since $\mathcal{I}(\alpha)$ varies on the same scale. They are genuine freezing points of the state, but shallow ones: flips in other bands, such as those of the edge atoms, are not suppressed at the same frequencies, and set the residual infidelity. In the families at $a=4.6\,\mu$m, individual configurations with a dominant band display their own combs, and the family average retains only the minima shared by many of them.

\paragraph{Family averages and the role of the drive.} Under single-frequency driving the low-frequency minima of the averages remain pinned close to the single-atom positions for all densities and spacings, set by the $k_2=0$ band, while the high-frequency minimum shifts and broadens as the spacing decreases. Under bi-frequency driving the averages lie systematically lower than under single-frequency driving at $a=5.4$ and $6\,\mu$m, and the advantage is lost at $a=4.6\,\mu$m, where the $k_2=2$ band approaches $\Delta_0$ and receives no suppression from the second harmonic [equation~\eqref{eq:Salpha}].

\paragraph{System size.} The positions of the minima of the family averages are insensitive to the system size. For $\rho=1/4$ at $a=5.4\,\mu$m we have compared the $L=12$ average with an average over a random sample of states with $n=4$ in a chain of $L=16$ atoms (appendix~\ref{app:size}). The two coincide up to a uniform offset $\ln(16/12)$, as expected from equation~\eqref{eq:F1}, according to which $1-F$ is extensive in the number of flippable atoms.

\subsection{Freezing landscape in the $(\omega,a)$ plane}
\label{sec:families:contour}

\begin{figure*}
{\centering
\includegraphics[width=\textwidth]{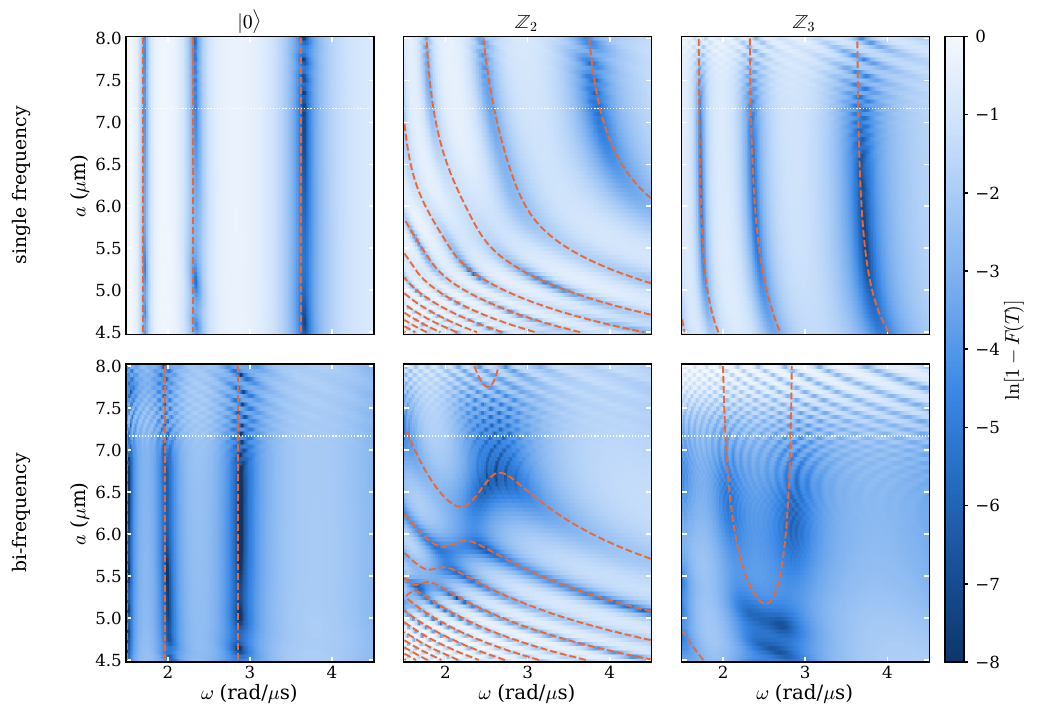}\par}
\caption{\label{fig:contour}Logarithmic infidelity $\ln[1-F(T)]$ in the $(\omega,a)$ plane for $\ket{0}$, $\ket{\mathbb{Z}_2}$ and $\ket{\mathbb{Z}_3}=\ket{010010\cdots010}$ (columns), for single-frequency (top) and bi-frequency (bottom) driving, computed in the full Hilbert space for $L=12$. Dashed lines: first-order freezing condition $\mathcal{I}(\alpha)=0$ for the bulk shift of each state. White dotted line: $V(a)=2\Delta_0$, above which the blockade fails during the sweep. The fine structure above this line consists of narrow resonances that are not fully resolved on the grid used.}
\end{figure*}

The four spacings considered so far sample the freezing landscape at isolated points. To follow its evolution continuously, and across the breakdown of the blockade, figure~\ref{fig:contour} shows the infidelity in the $(\omega,a)$ plane for the vacuum, $\ket{\mathbb{Z}_2}$ and $\ket{\mathbb{Z}_3}$, computed in the full Hilbert space so that no constraint needs to be switched off at large spacing. Since $\mathcal{I}(\alpha)=e^{i\pi\alpha/\omega}R(\alpha/\omega,x)$ with $R$ real [equation~\eqref{eq:I}], the first-order freezing condition $\mathcal{I}(\alpha)=0$ defines curves rather than isolated points in this plane. For each state we overlay these curves, evaluated for its bulk shift $\alpha(a)$ (table~\ref{tab:alpha}); they contain no free parameter.

\paragraph{Single-frequency driving.} For the vacuum the freezing lines are vertical: with $\alpha=0$ the condition reduces to $J_0(\Delta_0/\omega)=0$ at every spacing. For the density-wave states the dark valleys of the numerics follow the first-order curves over the whole range of spacings. For $\ket{\mathbb{Z}_3}$, whose shift is small, the curves bend only gently away from the vacuum positions as $a$ decreases. For $\ket{\mathbb{Z}_2}$ they bend strongly towards lower frequencies, and new branches enter from the low-frequency side each time $\alpha/\omega$ passes through another half-integer, so that the isolated minima at large spacing evolve continuously into the dense comb at small spacing (section~\ref{sec:families:n}). The figure thus makes explicit that the freezing frequencies are not fixed numbers but functions of the interaction shift, which the spacing tunes continuously through $\alpha\propto a^{-6}$.

\paragraph{Bi-frequency driving.} With the second harmonic the vacuum again freezes along vertical lines, now at the zeros of $2J_1(x)/x$, with low infidelity throughout the region between them. For $\ket{\mathbb{Z}_2}$ at small and intermediate spacing and for $\ket{\mathbb{Z}_3}$ at intermediate spacing the valleys follow the first-order curves as before. Two regions of strong freezing lie away from these curves: for $\ket{\mathbb{Z}_2}$ close to the vacuum freezing frequency at the upper end of the constrained regime, and for $\ket{\mathbb{Z}_3}$ at the smallest spacings, where its two first-order zeros approach each other and disappear. The dashed curves mark only exact zeros of the amplitude; in both regions $|\mathcal{I}(\alpha)|$ instead passes through a deep minimum without vanishing, which suppresses the dynamics almost as effectively. Along cuts through these regions, first-order perturbation theory including all available flips [equation~\eqref{eq:F1}] reproduces the exact infidelity of $\ket{\mathbb{Z}_2}$ quantitatively, while for $\ket{\mathbb{Z}_3}$ it captures the minimum but overestimates its depth, a discrepancy that is removed at second order.

\paragraph{Breakdown of the blockade.} Above the dotted line the nearest-neighbour coupling approaches the scale of the detuning sweep, and facilitation sets in (section~\ref{sec:fpt:cascade}). Under bi-frequency driving this region is filled with narrow resonances for all three states. They are strongest for $\ket{\mathbb{Z}_3}$, whose bulk contains facilitated atoms, weaker for $\ket{\mathbb{Z}_2}$, whose bulk atoms are dynamically blocked and whose resonances originate from the edge and from facilitated channels opened by de-excitations, and weakest for the vacuum, for which an excitation must be created first. For the vacuum this happens through the two-photon creation of a nearest-neighbour pair, whose resonance $2\Delta(t)=V(a)$ is crossed during the sweep precisely above the dotted line (section~\ref{sec:bifreq}). Under single-frequency driving the same resonances are present but faint, consistent with the weaker Rabi coupling at the turning points of the sweep.

\paragraph{The freezing window.} Taken together, the maps delineate a window of spacings within which freezing is both strong and controlled by the first-order condition. At its lower end the next-nearest-neighbour shift $2V(2a)$ approaches $\Delta_0$ and the freezing lines of states with excitations crowd into a comb; at its upper end the nearest-neighbour coupling becomes comparable to $\Delta_0$ and facilitation destroys the freezing of states with excitations. The same two edges bound the range in which the second harmonic is useful. Comparing the two rows of figure~\ref{fig:contour}, bi-frequency driving improves the freezing of the vacuum at every spacing, since none of its first-order flips carries a shift. For $\ket{\mathbb{Z}_2}$ the improvement is largest in the middle of the window and vanishes at both ends, where its relevant shift, $2V(2a)$ at small spacing and the facilitated nearest-neighbour coupling at large spacing, approaches $\Delta_0$. For $\ket{\mathbb{Z}_3}$, whose next-nearest-neighbour shift remains small, the improvement persists down to the smallest spacing but is lost, and eventually reversed, once its bulk atoms become facilitated. Each state thus benefits from the second harmonic exactly as long as all of its relevant shifts remain below $\Delta_0/\sqrt{2}$, as anticipated from equation~\eqref{eq:Salpha}. Both edges are set by the condition that an interaction shift reaches the detuning amplitude, and the non-monotonic dependence of the window-averaged infidelity on $a$ (figure~\ref{fig:summary}) is the projection of this window onto the spacing axis. Within the window, the freezing frequencies of any state follow from its local environments through $\mathcal{I}(\alpha)$, and the PXP limit, in which they would coincide with those of the vacuum, is almost never reached for the detuning amplitude used here.
\section{Discussion}
\label{sec:discussion}

We have studied how dynamical freezing in periodically driven Rydberg chains depends on the initial state and on the interatomic spacing, for single- and bi-frequency modulation of the drive. The central result is that, once the interaction tails are retained, the freezing condition is no longer a property of the drive alone but of each individual spin flip. In the interaction picture every flip carries a phase set by its interaction shift $\alpha_i$, the energy cost imposed by the excitations already present. To first order in the Rabi frequency the drive enters only through a single amplitude $\mathcal{I}(\alpha)$, which interpolates between the single-atom condition at $\alpha=0$ and a dense set of shifted zeros for $\alpha\gg\omega$. A product state freezes only where the amplitudes of all its available flips are suppressed simultaneously, so that its freezing landscape is fixed by the set of local environments it contains. This picture accounts for the behaviour of all the states we have considered: the robustness of the vacuum, whose first flips carry no shift; the continuous drift of the freezing frequencies of the density-wave states with spacing; the dense combs of narrow minima at small spacing; and the sorting of whole families of states by a single local motif.

Several features of this picture are specific to the van der Waals interaction. Because of the steep $r^{-6}$ decay, the shifts in a one-dimensional chain cluster into three bands set by the number of excitations two sites away, independently of the density, and the number of distinct environments therefore saturates rather than growing with the number of excitations. The same steepness fixes the ratio between next-nearest- and nearest-neighbour interactions, which leads to the main limitation we have identified. Freezing is strong and well described by the first-order condition only within a window of spacings bounded by two conditions of the same kind: at small spacing the next-nearest-neighbour shift reaches the detuning amplitude, and at large spacing the nearest-neighbour coupling does, at which point facilitation replaces the blockade. A PXP-like regime would require the blockade to hold throughout the sweep and the tails to be negligible on the scale of the drive frequency, $\Delta_0\ll V(a)\ll64\,\omega$, which is possible only for moderate $\Delta_0/\omega$. The regime of large $\Delta_0/\omega$, in which freezing points are numerous and the perturbative description is controlled, therefore leaves almost no room for the PXP limit, which serves as a useful reference rather than a regime that can be reached by tuning the spacing. We expect the window to be narrower in two-dimensional arrays, where the ratio between the first two interaction shells is much smaller than in a chain, consistent with the stronger geometry dependence observed in Ref.~\cite{Sarkar2026harmonic}.

The comparison between the two drive protocols follows from the same analysis. The second harmonic does not act on all flips equally. Its effect depends on the interaction shift of each flip [equation~\eqref{eq:Salpha}]. Flips with small shifts are strongly suppressed, those with intermediate shifts progressively less so, and flips whose shift approaches the detuning amplitude are enhanced rather than suppressed. Bi-frequency driving therefore improves freezing for every initial state whose available flips all have small shifts. The vacuum, whose flips carry no shift, is the clearest example. The advantage is lost, or even reversed, for states that contain flips close to the facilitation condition. It provides a simple design criterion: harmonic engineering of the drive is effective as long as the interaction shifts of the states of interest remain below the threshold. More generally, since the freezing lines in the $(\omega,a)$ plane are known analytically, the spacing and drive frequency can be chosen to freeze a targeted class of configurations while leaving others to evolve, which suggests a route towards state-selective control in programmable arrays.

Our results are directly accessible in current neutral-atom platforms~\cite{Bernien2017,Bluvstein2021,Wurtz2023}, where the spacing, the detuning and the Rabi frequency are all independently programmable and product states with arbitrary excitation patterns can be prepared. The one-period fidelities we have studied correspond to the stroboscopic observables measured in Ref.~\cite{Sarkar2026harmonic}, and the predicted dependence of the freezing frequencies on spacing and on the arrangement of the excitations could be tested with modest modifications of existing protocols. Our analysis is restricted to a single drive period, to product initial states, and to system sizes accessible to exact numerics; the persistence of the environment-selective freezing over many periods, where higher-order processes and heating eventually dominate, remains an open question. Initial states that are superpositions of configurations with different environments, for which the different flip classes can interfere, will be addressed in future work.

\begin{acknowledgements}
M.S. and S.B acknowledge EPSRC-SFI funded project EP/X039889/1
(GeQuantumBus) for support. 
\end{acknowledgements}

\section*{Data availability statement}
The data that support the findings of this study are available upon reasonable request from the authors.

\section*{References}
\bibliography{refs}

@article{Bukov2015,
  author  = {Bukov, M. and D'Alessio, L. and Polkovnikov, A.},
  title   = {Universal high-frequency behavior of periodically driven systems: from dynamical stabilization to {F}loquet engineering},
  journal = {Adv. Phys.}, volume = {64}, pages = {139}, year = {2015},
  url     = {https://doi.org/10.1080/00018732.2015.1055918}}

@article{Eckardt2017,
  author  = {Eckardt, A.},
  title   = {Colloquium: Atomic quantum gases in periodically driven optical lattices},
  journal = {Rev. Mod. Phys.}, volume = {89}, pages = {011004}, year = {2017},
  url     = {https://doi.org/10.1103/RevModPhys.89.011004}}

@article{Oka2019,
  author  = {Oka, T. and Kitamura, S.},
  title   = {{F}loquet engineering of quantum materials},
  journal = {Annu. Rev. Condens. Matter Phys.}, volume = {10}, pages = {387}, year = {2019},
  url     = {https://doi.org/10.1146/annurev-conmatphys-031218-013423}}

@article{Sen2021JPCM,
  author  = {Sen, A. and Sen, D. and Sengupta, K.},
  title   = {Analytic approaches to periodically driven closed quantum systems: methods and applications},
  journal = {J. Phys.: Condens. Matter}, volume = {33}, pages = {443003}, year = {2021},
  url     = {https://doi.org/10.1088/1361-648X/ac1b61}}

@article{Zhang2017DTC,
  author  = {Zhang, J. and others},
  title   = {Observation of a discrete time crystal},
  journal = {Nature}, volume = {543}, pages = {217}, year = {2017},
  url     = {https://doi.org/10.1038/nature21413}}

@article{Choi2017DTC,
  author  = {Choi, S. and others},
  title   = {Observation of discrete time-crystalline order in a disordered dipolar many-body system},
  journal = {Nature}, volume = {543}, pages = {221}, year = {2017},
  url     = {https://doi.org/10.1038/nature21426}}

@article{Mi2022DTC,
  author  = {Mi, X. and others},
  title   = {Time-crystalline eigenstate order on a quantum processor},
  journal = {Nature}, volume = {601}, pages = {531}, year = {2022},
  url     = {https://doi.org/10.1038/s41586-021-04257-w}}

@article{Dumitrescu2022,
  author  = {Dumitrescu, P. T. and others},
  title   = {Dynamical topological phase realized in a trapped-ion quantum simulator},
  journal = {Nature}, volume = {607}, pages = {463}, year = {2022},
  url     = {https://doi.org/10.1038/s41586-022-04853-4}}

@article{DAlessio2014,
  author  = {D'Alessio, L. and Rigol, M.},
  title   = {Long-time behavior of isolated periodically driven interacting lattice systems},
  journal = {Phys. Rev. X}, volume = {4}, pages = {041048}, year = {2014},
  url     = {https://doi.org/10.1103/PhysRevX.4.041048}}

@article{Lazarides2014,
  author  = {Lazarides, A. and Das, A. and Moessner, R.},
  title   = {Equilibrium states of generic quantum systems subject to periodic driving},
  journal = {Phys. Rev. E}, volume = {90}, pages = {012110}, year = {2014},
  url     = {https://doi.org/10.1103/PhysRevE.90.012110}}

@article{Ponte2015,
  author  = {Ponte, P. and Chandran, A. and Papi\'c, Z. and Abanin, D. A.},
  title   = {Periodically driven ergodic and many-body localized quantum systems},
  journal = {Ann. Phys.}, volume = {353}, pages = {196}, year = {2015},
  url     = {https://doi.org/10.1016/j.aop.2014.11.008}}

@article{Abanin2015,
  author  = {Abanin, D. A. and De Roeck, W. and Huveneers, F.},
  title   = {Exponentially slow heating in periodically driven many-body systems},
  journal = {Phys. Rev. Lett.}, volume = {115}, pages = {256803}, year = {2015},
  url     = {https://doi.org/10.1103/PhysRevLett.115.256803}}

@article{Mori2016,
  author  = {Mori, T. and Kuwahara, T. and Saito, K.},
  title   = {Rigorous bound on energy absorption and generic relaxation in periodically driven quantum systems},
  journal = {Phys. Rev. Lett.}, volume = {116}, pages = {120401}, year = {2016},
  url     = {https://doi.org/10.1103/PhysRevLett.116.120401}}

@article{Kuwahara2016,
  author  = {Kuwahara, T. and Mori, T. and Saito, K.},
  title   = {{F}loquet--{M}agnus theory and generic transient dynamics in periodically driven many-body quantum systems},
  journal = {Ann. Phys.}, volume = {367}, pages = {96}, year = {2016},
  url     = {https://doi.org/10.1016/j.aop.2016.01.012}}

@article{Peng2021,
  author  = {Peng, P. and Yin, C. and Huang, X. and Ramanathan, C. and Cappellaro, P.},
  title   = {{F}loquet prethermalization in dipolar spin chains},
  journal = {Nat. Phys.}, volume = {17}, pages = {444}, year = {2021},
  url     = {https://doi.org/10.1038/s41567-020-01120-z}}

@article{Else2020,
  author  = {Else, D. V. and Ho, W. W. and Quito, P. T.},
  title   = {Long-lived interacting phases of matter protected by multiple time-translation symmetries in quasiperiodically driven systems},
  journal = {Phys. Rev. X}, volume = {10}, pages = {021032}, year = {2020},
  url     = {https://doi.org/10.1103/PhysRevX.10.021032}}

@article{Dunlap1986,
  author  = {Dunlap, D. H. and Kenkre, V. M.},
  title   = {Dynamic localization of a charged particle moving under the influence of an electric field},
  journal = {Phys. Rev. B}, volume = {34}, pages = {3625}, year = {1986},
  url     = {https://doi.org/10.1103/PhysRevB.34.3625}}

@article{Grossmann1991,
  author  = {Grossmann, F. and Dittrich, T. and Jung, P. and H\"anggi, P.},
  title   = {Coherent destruction of tunneling},
  journal = {Phys. Rev. Lett.}, volume = {67}, pages = {516}, year = {1991},
  url     = {https://doi.org/10.1103/PhysRevLett.67.516}}

@article{Lignier2007,
  author  = {Lignier, H. and Sias, C. and Ciampini, D. and Singh, Y. and Zenesini, A. and Morsch, O. and Arimondo, E.},
  title   = {Dynamical control of matter-wave tunneling in periodic potentials},
  journal = {Phys. Rev. Lett.}, volume = {99}, pages = {220403}, year = {2007},
  url     = {https://doi.org/10.1103/PhysRevLett.99.220403}}

@article{Das2010,
  author  = {Das, A.},
  title   = {Exotic freezing of response in a quantum many-body system},
  journal = {Phys. Rev. B}, volume = {82}, pages = {172402}, year = {2010},
  url     = {https://doi.org/10.1103/PhysRevB.82.172402}}

@article{Bhattacharyya2012,
  author  = {Bhattacharyya, S. and Das, A. and Dasgupta, S.},
  title   = {Transverse {I}sing chain under periodic instantaneous quenches: Dynamical many-body freezing and emergence of slow solitary oscillations},
  journal = {Phys. Rev. B}, volume = {86}, pages = {054410}, year = {2012},
  url     = {https://doi.org/10.1103/PhysRevB.86.054410}}

@article{Hegde2014,
  author  = {Hegde, S. S. and Katiyar, H. and Mahesh, T. S. and Das, A.},
  title   = {Freezing a quantum magnet by repeated quantum interference: An experimental realization },
  journal = {Phys. Rev. B}, volume = {90}, pages = {174407}, year = {2014},
  url     = {https://doi.org/10.1103/PhysRevB.90.174407}}

@article{Haldar2021,
  author  = {Haldar, A. and Sen, D. and Moessner, R. and Das, A.},
  title   = {Dynamical freezing and scar points in strongly driven {F}loquet matter: Resonance vs emergent conservation laws},
  journal = {Phys. Rev. X}, volume = {11}, pages = {021008}, year = {2021},
  url     = {https://doi.org/10.1103/PhysRevX.11.021008}}

@article{Haldar2024,
  author  = {Haldar, A. and Das, A. and Chaudhuri, S. and Staszewski, L. and Wietek, A. and Pollmann, F. and Moessner, R. and Das, A.},
  title   = {Dynamical freezing in the thermodynamic limit: the strongly driven ensemble},
  journal = {arXiv:2410.11050}, year = {2024},
  url     = {https://arxiv.org/abs/2410.11050}}

@article{Kar2016,
  author  = {Kar, S. and Mukherjee, B. and Sengupta, K.},
  title   = {Tuning towards dynamic freezing using a two-rate protocol},
  journal = {Phys. Rev. B}, volume = {94}, pages = {075130}, year = {2016},
  url     = {https://doi.org/10.1103/PhysRevB.94.075130}}

@article{Ghosh2018,
  author  = {Ghosh, R. and Sen, A. and Sengupta, K.},
  title   = {Ramp and periodic dynamics across non-{I}sing critical points},
  journal = {Phys. Rev. B}, volume = {97}, pages = {014309}, year = {2018},
  url     = {https://doi.org/10.1103/PhysRevB.97.014309}}

@article{Mukherjee2020scars,
  author  = {Mukherjee, B. and Nandy, S. and Sen, A. and Sen, D. and Sengupta, K.},
  title   = {Collapse and revival of quantum many-body scars via {F}loquet engineering},
  journal = {Phys. Rev. B}, volume = {101}, pages = {245107}, year = {2020},
  url     = {https://doi.org/10.1103/PhysRevB.101.245107}}

@article{Mukherjee2020vacuum,
  author  = {Mukherjee, B. and Sen, A. and Sen, D. and Sengupta, K.},
  title   = {Dynamics of the vacuum state in a periodically driven {R}ydberg chain},
  journal = {Phys. Rev. B}, volume = {102}, pages = {075123}, year = {2020},
  url     = {https://doi.org/10.1103/PhysRevB.102.075123}}

@article{Guo2025,
  author  = {Guo, H. and Mukherjee, R. and Chowdhury, D.},
  title   = {Dynamical freezing in exactly solvable models of driven chaotic quantum dots},
  journal = {Phys. Rev. Lett.}, volume = {134}, pages = {226501}, year = {2025},
  url     = {https://doi.org/10.1103/ggk3-6cf8}}

@article{Mukherjee2026instanton,
  author  = {Mukherjee, R. and Guo, H. and Chowdhury, D.},
  title   = {{F}loquet thermalization via instantons near dynamical freezing},
  journal = {Phys. Rev. X}, volume = {16}, pages = {011041}, year = {2026},
  url     = {https://doi.org/10.1103/4w5w-57my}}

@article{Ghosh2020FPT,
  author  = {Ghosh, R. and Mukherjee, B. and Sengupta, K.},
  title   = {{F}loquet perturbation theory for periodically driven weakly interacting fermions},
  journal = {Phys. Rev. B}, volume = {102}, pages = {235114}, year = {2020},
  url     = {https://doi.org/10.1103/PhysRevB.102.235114}}

@misc{Sarkar2026harmonic,
      title={Harmonic Control of Dynamical Freezing in Programmable {R}ydberg Atom Arrays}, 
      author={Madhumita Sarkar and Ben Zindorf and Bhaskar Mukherjee and Sougato Bose and Roopayan Ghosh},
      year={2025},
      eprint={2511.09633},
      archivePrefix={arXiv},
      primaryClass={quant-ph},
      url={https://arxiv.org/abs/2511.09633}, 
}

@article{Saffman2010,
  author  = {Saffman, M. and Walker, T. G. and M{\o}lmer, K.},
  title   = {Quantum information with {R}ydberg atoms},
  journal = {Rev. Mod. Phys.}, volume = {82}, pages = {2313}, year = {2010},
  url     = {https://doi.org/10.1103/RevModPhys.82.2313}}

@article{Browaeys2020,
  author  = {Browaeys, A. and Lahaye, T.},
  title   = {Many-body physics with individually controlled {R}ydberg atoms},
  journal = {Nat. Phys.}, volume = {16}, pages = {132}, year = {2020},
  url     = {https://doi.org/10.1038/s41567-019-0733-z}}

@article{Bernien2017,
  author  = {Bernien, H. and others},
  title   = {Probing many-body dynamics on a 51-atom quantum simulator},
  journal = {Nature}, volume = {551}, pages = {579}, year = {2017},
  url     = {https://doi.org/10.1038/nature24622}}

@article{Turner2018,
  author  = {Turner, C. J. and Michailidis, A. A. and Abanin, D. A. and Serbyn, M. and Papi\'c, Z.},
  title   = {Weak ergodicity breaking from quantum many-body scars},
  journal = {Nat. Phys.}, volume = {14}, pages = {745}, year = {2018},
  url     = {https://doi.org/10.1038/s41567-018-0137-5}}

@article{Bluvstein2021,
  author  = {Bluvstein, D. and others},
  title   = {Controlling quantum many-body dynamics in driven {R}ydberg atom arrays},
  journal = {Science}, volume = {371}, pages = {1355}, year = {2021},
  url     = {https://doi.org/10.1126/science.abg2530}}

@article{Ebadi2021,
  author  = {Ebadi, S. and others},
  title   = {Quantum phases of matter on a 256-atom programmable quantum simulator},
  journal = {Nature}, volume = {595}, pages = {227}, year = {2021},
  url     = {https://doi.org/10.1038/s41586-021-03582-4}}

@article{Semeghini2021,
  author  = {Semeghini, G. and others},
  title   = {Probing topological spin liquids on a programmable quantum simulator},
  journal = {Science}, volume = {374}, pages = {1242}, year = {2021},
  url     = {https://doi.org/10.1126/science.abi8794}}

@article{Keesling2019,
  author  = {Keesling, A. and others},
  title   = {Quantum {K}ibble--{Z}urek mechanism and critical dynamics on a programmable {R}ydberg simulator},
  journal = {Nature}, volume = {568}, pages = {207}, year = {2019},
  url     = {https://doi.org/10.1038/s41586-019-1070-1}}

@article{Darbha2025,
  author  = {Darbha, S. and others},
  title   = {Probing emergent prethermal dynamics and resonant melting on a programmable quantum simulator},
  journal = {arXiv:2510.11706}, year = {2025},
  url     = {https://arxiv.org/abs/2510.11706}}

@article{Wurtz2023,
  author  = {Wurtz, J. and others},
  title   = {Aquila: {Q}u{E}ra's 256-qubit neutral-atom quantum computer},
  journal = {arXiv:2306.11727}, year = {2023},
  url     = {https://arxiv.org/abs/2306.11727}}

@article{Fendley2004,
  author  = {Fendley, P. and Sengupta, K. and Sachdev, S.},
  title   = {Competing density-wave orders in a one-dimensional hard-boson model},
  journal = {Phys. Rev. B}, volume = {69}, pages = {075106}, year = {2004},
  url     = {https://doi.org/10.1103/PhysRevB.69.075106}}

@article{Lesanovsky2012,
  author  = {Lesanovsky, I. and Katsura, H.},
  title   = {Interacting {F}ibonacci anyons in a {R}ydberg gas},
  journal = {Phys. Rev. A}, volume = {86}, pages = {041601(R)}, year = {2012},
  url     = {https://doi.org/10.1103/PhysRevA.86.041601}}

@article{Lu2026,
  title = {Dynamical freezing for magnetometry in an interacting spin ensemble},
  volume = {653},
  issn = {1476-4687},
  url = {http://dx.doi.org/10.1038/s41586-026-10585-6},
  doi = {10.1038/s41586-026-10585-6},
  number = {8116},
  journal = {Nature},
  publisher = {Springer Science and Business Media LLC},
  author = {Lu, Ya-Nan and Yuan, Dong and Ma, Yixuan and Liu, Yan-Qing and Jiang, Si and Meng, Xiang-Qian and Xu, Yi-Jie and Chang, Xiu-Ying and Zu, Chong and Zhao, Hong-Zheng and Deng, Dong-Ling and Duan, Lu-Ming and Hou, Pan-Yu},
  year = {2026},
  month = May,
  pages = {1027–1032},
}

@article{Sarkar2023ladder,
  author  = {Sarkar, M. and Pal, M. and Sen, A. and Sengupta, K.},
  title   = {Quantum order-by-disorder induced phase transition in {R}ydberg ladders with staggered detuning},
  journal = {SciPost Phys.}, volume = {14}, pages = {004}, year = {2023},
  doi     = {10.21468/SciPostPhys.14.1.004},
  url     = {https://doi.org/10.21468/SciPostPhys.14.1.004}}

@article{Pal2025scar,
  author  = {Pal, M. and Sarkar, M. and Sengupta, K. and Sen, A.},
  title   = {Scar-induced imbalance in staggered {R}ydberg ladders},
  journal = {Phys. Rev. B}, volume = {111}, pages = {L161101}, year = {2025},
  doi     = {10.1103/PhysRevB.111.L161101},
  url     = {https://doi.org/10.1103/PhysRevB.111.L161101}}

@article{Turner2018PRB,
  author  = {Turner, C. J. and Michailidis, A. A. and Abanin, D. A. and Serbyn, M. and Papi\'c, Z.},
  title   = {Quantum scarred eigenstates in a {R}ydberg atom chain: Entanglement, breakdown of thermalization, and stability to perturbations},
  journal = {Phys. Rev. B}, volume = {98}, pages = {155134}, year = {2018},
  doi     = {10.1103/PhysRevB.98.155134},
  url     = {https://doi.org/10.1103/PhysRevB.98.155134}}

@article{Ho2019,
  author  = {Ho, W. W. and Choi, S. and Pichler, H. and Lukin, M. D.},
  title   = {Periodic orbits, entanglement, and quantum many-body scars in constrained models: Matrix product state approach},
  journal = {Phys. Rev. Lett.}, volume = {122}, pages = {040603}, year = {2019},
  doi     = {10.1103/PhysRevLett.122.040603},
  url     = {https://doi.org/10.1103/PhysRevLett.122.040603}}

@article{Khemani2019,
  author  = {Khemani, V. and Laumann, C. R. and Chandran, A.},
  title   = {Signatures of integrability in the dynamics of {R}ydberg-blockaded chains},
  journal = {Phys. Rev. B}, volume = {99}, pages = {161101(R)}, year = {2019},
  doi     = {10.1103/PhysRevB.99.161101},
  url     = {https://doi.org/10.1103/PhysRevB.99.161101}}

@article{Lin2019,
  author  = {Lin, C.-J. and Motrunich, O. I.},
  title   = {Exact quantum many-body scar states in the {R}ydberg-blockaded atom chain},
  journal = {Phys. Rev. Lett.}, volume = {122}, pages = {173401}, year = {2019},
  doi     = {10.1103/PhysRevLett.122.173401},
  url     = {https://doi.org/10.1103/PhysRevLett.122.173401}}

@article{Choi2019,
  author  = {Choi, S. and Turner, C. J. and Pichler, H. and Ho, W. W. and Michailidis, A. A. and Papi\'c, Z. and Serbyn, M. and Lukin, M. D. and Abanin, D. A.},
  title   = {Emergent {SU}(2) dynamics and perfect quantum many-body scars},
  journal = {Phys. Rev. Lett.}, volume = {122}, pages = {220603}, year = {2019},
  doi     = {10.1103/PhysRevLett.122.220603},
  url     = {https://doi.org/10.1103/PhysRevLett.122.220603}}

@article{Serbyn2021,
  author  = {Serbyn, M. and Abanin, D. A. and Papi\'c, Z.},
  title   = {Quantum many-body scars and weak breaking of ergodicity},
  journal = {Nat. Phys.}, volume = {17}, pages = {675}, year = {2021},
  doi     = {10.1038/s41567-021-01230-2},
  url     = {https://doi.org/10.1038/s41567-021-01230-2}}

@article{Maskara2021,
  author  = {Maskara, N. and Michailidis, A. A. and Ho, W. W. and Bluvstein, D. and Choi, S. and Lukin, M. D. and Serbyn, M.},
  title   = {Discrete time-crystalline order enabled by quantum many-body scars: Entanglement steering via periodic driving},
  journal = {Phys. Rev. Lett.}, volume = {127}, pages = {090602}, year = {2021},
  doi     = {10.1103/PhysRevLett.127.090602},
  url     = {https://doi.org/10.1103/PhysRevLett.127.090602}}

@article{Hudomal2022,
  author  = {Hudomal, A. and Desaules, J.-Y. and Mukherjee, B. and Su, G.-X. and Halimeh, J. C. and Papi\'c, Z.},
  title   = {Driving quantum many-body scars in the {PXP} model},
  journal = {Phys. Rev. B}, volume = {106}, pages = {104302}, year = {2022},
  doi     = {10.1103/PhysRevB.106.104302},
  url     = {https://doi.org/10.1103/PhysRevB.106.104302}}

@article{Chepiga2021,
  author  = {Chepiga, N. and Mila, F.},
  title   = {Kibble--{Z}urek exponent and chiral transition of the period-4 phase of {R}ydberg chains},
  journal = {Nat. Commun.}, volume = {12}, pages = {414}, year = {2021},
  doi     = {10.1038/s41467-020-20641-y},
  url     = {https://doi.org/10.1038/s41467-020-20641-y}}

@article{Samajdar2020,
  author  = {Samajdar, R. and Ho, W. W. and Pichler, H. and Lukin, M. D. and Sachdev, S.},
  title   = {Complex density wave orders and quantum phase transitions in a model of square-lattice {R}ydberg atom arrays},
  journal = {Phys. Rev. Lett.}, volume = {124}, pages = {103601}, year = {2020},
  doi     = {10.1103/PhysRevLett.124.103601},
  url     = {https://doi.org/10.1103/PhysRevLett.124.103601}}

@article{Verresen2021,
  author  = {Verresen, R. and Lukin, M. D. and Vishwanath, A.},
  title   = {Prediction of toric code topological order from {R}ydberg blockade},
  journal = {Phys. Rev. X}, volume = {11}, pages = {031005}, year = {2021},
  doi     = {10.1103/PhysRevX.11.031005},
  url     = {https://doi.org/10.1103/PhysRevX.11.031005}}

@article{Banerjee2025flatband,
  author  = {Banerjee, Tista and Choudhury, Sayan and Sengupta, K.},
  title   = {Exact {F}loquet flat band and heating suppression via two-tone drive protocols},
  journal = {New J. Phys.},
  volume  = {27},
  pages   = {084506},
  year    = {2025},
  doi     = {10.1088/1367-2630/adfd07},
  url     = {https://doi.org/10.1088/1367-2630/adfd07}
}

@article{Ghosh2025tworate,
  author  = {Ghosh, Krishanu and Choudhury, Sayan and Sen, Diptiman and Sengupta, K.},
  title   = {Heating suppression via two-rate random and quasiperiodic drive protocols},
  journal = {Phys. Rev. B},
  volume  = {112},
  pages   = {155142},
  year    = {2025},
  doi     = {10.1103/3slf-yvtq},
  url     = {https://doi.org/10.1103/3slf-yvtq}
}

@misc{Banerjee2026pairproduction,
  author        = {Banerjee, Tista and Sengupta, K.},
  title         = {Resonant pair production and flat band melting in a driven {R}ydberg chain},
  year          = {2026},
  eprint        = {2610.22532},
  archivePrefix = {arXiv},
  primaryClass  = {quant-ph},
  doi           = {10.48550/arXiv.2610.22532},
  url           = {https://arxiv.org/abs/2610.22532}
}
\onecolumngrid
\appendix

\section{Floquet perturbation theory}
\label{app:fpt}

\subsection{First-order amplitude}
Inserting the expansion~\eqref{eq:Bn} into $H_F^{(1)}=T^{-1}\int_0^TH_I(t)\,dt$, each flip contributes
\begin{equation}
\mathcal{I}(\alpha)=\sum_nB_n(x)\,\frac{1}{T}\int_0^Tdt\,e^{i(\alpha-n\omega)t}
=\sum_nB_n(x)\,e^{i\pi(u-n)}\frac{\sin\pi(u-n)}{\pi(u-n)},\qquad u=\frac{\alpha}{\omega}.
\end{equation}
Since $e^{-i\pi n}\sin\pi(u-n)=\sin\pi u$, the $n$-dependence of the prefactor drops out and equation~\eqref{eq:I} follows. For integer $u=m$ only the term $n=m$ survives and $\mathcal{I}=B_m(x)$. Because $B_n$ is real, $\mathcal{I}(\alpha)=e^{i\pi u}R(u,x)$ with $R$ real, so that the freezing condition $R=0$ defines curves in the $(\omega,a)$ plane.

\subsection{Bi-frequency coefficients}
Applying the recurrence $J_{m-1}(x)+J_{m+1}(x)=(2m/x)J_m(x)$ for $m=n\pm1$ and adding the two relations gives $J_{n-2}+J_{n+2}+2J_n=(2/x)[(n-1)J_{n-1}+(n+1)J_{n+1}]$, so that $B_n=x^{-1}[(n-1)J_{n-1}+(n+1)J_{n+1}]$. Writing $(n-1)J_{n-1}+(n+1)J_{n+1}=n(J_{n-1}+J_{n+1})-(J_{n-1}-J_{n+1})$ and using $J_{n-1}+J_{n+1}=(2n/x)J_n$ and $J_{n-1}-J_{n+1}=2J_n'$ yields equation~\eqref{eq:Bnbi}. In the oscillatory regime $|n|<x$, with $|n|=x\cos\beta$, the Debye asymptotic forms read
\begin{equation}
J_n(x)\simeq\mathcal{A}\cos\varphi,\qquad J_n'(x)\simeq-\mathcal{A}\sin\beta\,\sin\varphi,\qquad\mathcal{A}=\sqrt{\frac{2}{\pi x\sin\beta}},
\end{equation}
with $\varphi=x\sin\beta-|n|\beta-\pi/4$. Inserting them into equation~\eqref{eq:Bnbi},
\begin{equation}
B_n(x)\simeq\frac{2\mathcal{A}}{x}\left[\frac{n^2}{x}\cos\varphi+\sin\beta\,\sin\varphi\right],
\end{equation}
whose envelope is smaller than that of $J_n$ by the factor $(2/x^2)\sqrt{n^4+x^2-n^2}$. With $n=\alpha/\omega$ and $x=\Delta_0/\omega$ this gives equation~\eqref{eq:Salpha}. For $x\approx8$-$12$ the asymptotic ratio agrees with the exact root-mean-square ratio $\langle B_n^2\rangle^{1/2}/\langle J_n^2\rangle^{1/2}$ to within about $10-20\%$, with the largest deviations close to $|n|=x$, where the Debye forms break down and the transition to the evanescent regime $|n|>x$, in which both coefficients are exponentially small, takes place.
\subsection{Second-order kernel}
\label{app:kernel}
The time ordering in equation~\eqref{eq:K} reflects the fact that the flip of atom $j$ at the earlier time $t_2$ changes the environment of atom $i$, so that the later flip at $t_1$ carries the shift $\alpha_i+s'V_{ij}$ evaluated in the initial configuration. Equivalently, moving all occupation-dependent exponentials to the left, the commutation phase is acquired at the time of the exponential being moved, $t_2$, and the shifts are evaluated in the final configuration. With $A=s(\alpha_i+s'V_{ij})-m\omega$ and $B=s'\alpha_j-n\omega$, the time integral is
\begin{equation}
\int_0^T\!dt_1\,e^{iAt_1}\int_0^{t_1}\!dt_2\,e^{iBt_2}=-\left[\frac{e^{i(A+B)T}-1}{B(A+B)}-\frac{e^{iAT}-1}{AB}\right]
\end{equation}
for $A,B,A+B\neq0$. The resonant cases follow as continuous limits: $[e^{iAT}(1-iAT)-1]/A^2$ for $B=0$; $(1-e^{iBT}+iBT)/B^2$ for $A=0$; $(1-e^{iAT}+iAT)/A^2$ for $A+B=0$; and $T^2/2$ for $A=B=0$. The last, disconnected, contribution is cancelled exactly by the $T^2\mathcal{I}\mathcal{I}/2$ term in equation~\eqref{eq:HF2}.

\subsection{Implementation and accuracy}
\begin{figure}
\centering
\includegraphics[width=0.48\columnwidth]{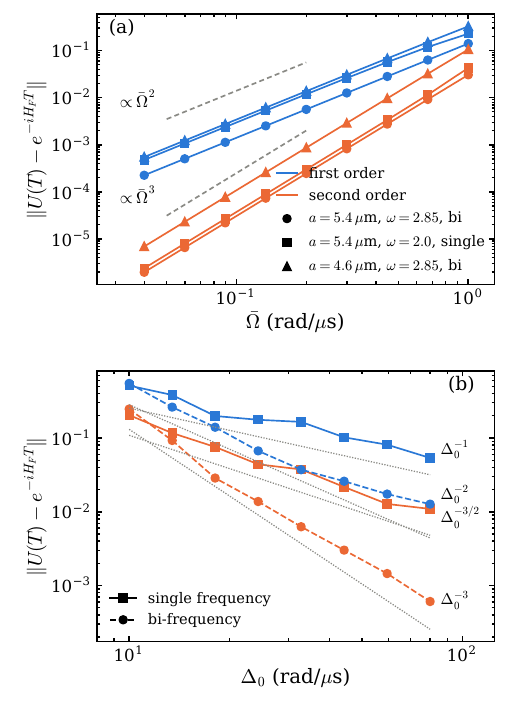}
\caption{\label{fig:appA}Operator-norm distance between the exact one-period evolution operator and $e^{-iH_FT}$ at first (blue) and second (orange) order. (a) As a function of $\bar\Omega$ at fixed $\omega$, for three representative cases ($L=8$, constrained space). (b) As a function of $\Delta_0$ at fixed $\bar\Omega=1\,\mathrm{rad}/\mu$s and $a=6\,\mu$m, for single-frequency (solid) and bi-frequency (dashed) driving; each point is a geometric mean over six drive frequencies. Dotted lines indicate the power laws discussed in the text.}
\end{figure}

The harmonic sums are truncated at $|n|\leq24$, which is sufficient over our frequency window, where $x=\Delta_0/\omega\lesssim13$. Since the kernel depends on the configuration only through the triple $(\alpha_i,\alpha_j,V_{ij})$, it is evaluated once for each distinct triple, which reduces the cost by an order of magnitude in the full Hilbert space. The resulting $H_F$ is Hermitian to machine precision. Figure~\ref{fig:appA} compares $e^{-iH_FT}$ with the exact one-period evolution operator: the error scales as $\bar\Omega^2$ at first order and as $\bar\Omega^3$ at second order, for constant $\Delta_0$ and $\omega$ as required. Panel (b) varies the detuning amplitude at fixed $\bar\Omega$ and $a$: for single-frequency driving the errors decay as $\Delta_0^{-1}$ and $\Delta_0^{-3/2}$, the second and third powers of $\bar\Omega/\sqrt{\Delta_0\omega}$, whereas for bi-frequency driving they decay as $\Delta_0^{-2}$ and $\Delta_0^{-3}$, the second and third powers of $\bar\Omega/\Delta_0$.

\section{Numerical method}
\label{app:numerics}

We integrate the Schr\"odinger equation in the interaction picture with respect to the diagonal part of the Hamiltonian, in which the state $\ket{\varphi(t)}=U_0^\dagger(t)\ket{\psi(t)}$ evolves under
\begin{equation}
\bra{c}H_I(t)\ket{c'}=\frac{\bar\Omega}{2}f(t)\,\bra{c}\sigma^x\ket{c'}\,e^{i[\Theta_c(t)-\Theta_{c'}(t)]},\qquad
\Theta_c(t)=-\frac{\Delta_0}{\omega}\sin(\omega t)\,N_c+V_c\,t,
\end{equation}
where $N_c$ and $V_c$ are the number of excitations and the interaction energy of configuration $c$, and $c,c'$ run over the configurations of the Hilbert space used, either the constrained or the full one. Since $U_0(T)$ is diagonal, the return probability of a product state is the same in both frames. All interaction energies, including the nearest-neighbour coupling in the full Hilbert space, enter only through phases, so that the time step is limited by the Rabi frequency and by the rate at which these phases wind, rather than by the largest energy in the problem. We use a fixed-step fourth-order Runge--Kutta scheme. The detuning phase depends on $c$ only through $N_c$ and is taken from a table of $L+1$ entries at each time, while $e^{iV_ct}$ is advanced by multiplication with a precomputed factor at every half step. All initial states of interest are propagated together, and different drive frequencies are distributed over parallel threads. The largest Hilbert spaces are $F_{23}=28657$ states for $L=21$ in the constrained space and $2^{16}$ states for $L=16$ in the full space.

\begin{figure}
\centering
\includegraphics[width=0.48\columnwidth]{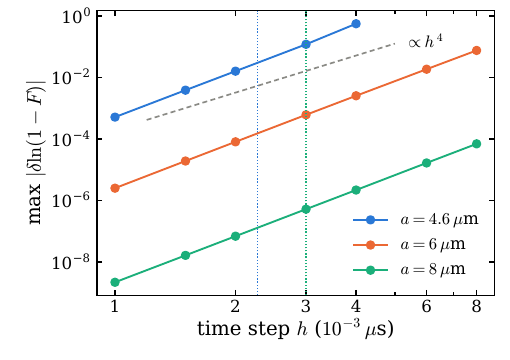}
\caption{\label{fig:appB}Maximal error in $\ln(1-F)$ over the drive window and the states $\ket{0}$, $\ket{\mathbb{Z}_2}$, $\ket{\mathbb{Z}_3}$, as a function of the time step, for bi-frequency driving in the full Hilbert space ($L=10$), relative to a reference with $h=2.5\times10^{-4}\,\mu$s. Vertical lines mark the step used for each spacing.}
\end{figure}

In the constrained space, the phases wind at most at the rate of the interaction tails, and a step $h=3\times10^{-3}\,\mu$s converges $\ln(1-F)$ to better than $10^{-5}$. In the full space, the nearest-neighbour coupling enters the phases and the error grows rapidly as the spacing decreases (figure~\ref{fig:appB}). We therefore use $h=\min[3\times10^{-3}\,\mu\mathrm{s},\,1.3/V(a)]$, which keeps the error below $10^{-3}$ for $a\geq6\,\mu$m and below a few times $10^{-2}$ at the smallest spacings of figure~\ref{fig:contour}, well below the resolution of the colour scale.

\section{Constrained versus full Hilbert space}
\label{app:blockade}

For $a\leq6\,\mu$m we have compared the constrained dynamics with simulations in the full Hilbert space for $L=10$, for all four ordered states and both protocols. The two differ by less than $5\times10^{-3}$ in $F$ over the entire frequency window, which justifies the use of the constrained space for the line plots and density families at these spacings.

\begin{figure*}
\centering
\includegraphics[width=\textwidth]{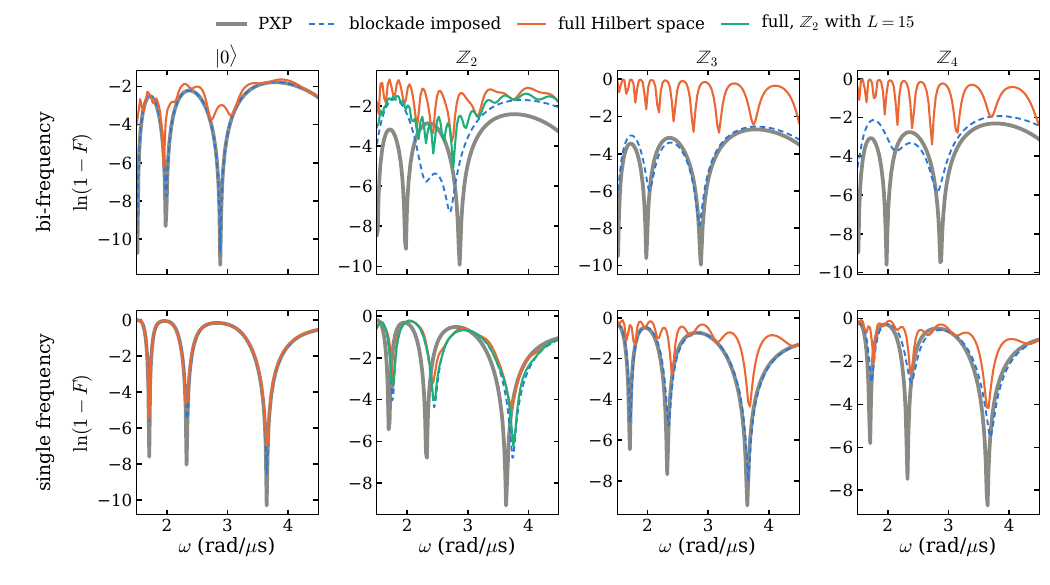}
\caption{\label{fig:appC}Infidelity at $a=8\,\mu$m for bi-frequency (top) and single-frequency (bottom) driving: PXP model (grey), blockade-constrained dynamics with interaction tails (blue dashed) and full Hilbert space (orange), for $L=16$ ($L=15$ for $\ket{\mathbb{Z}_3}$). Green: full Hilbert space for $\ket{\mathbb{Z}_2}$ with $L=15$, for which both end atoms are excited. The full-space curves for $L\geq14$ are computed on a frequency grid of spacing $0.03\,\mathrm{rad}/\mu$s.}
\end{figure*}

At $a=8\,\mu$m this is no longer the case. Figure~\ref{fig:appC} compares, for this spacing, the PXP model, the constrained dynamics with the tails retained, and the full Hilbert space. The constrained results lie close to the PXP limit, since the next-nearest-neighbour tails are weak, whereas the full-space results differ qualitatively for all states with excitations, in particular under bi-frequency driving, where facilitation replaces the deep PXP minima by a comb of shallow resonances. The vacuum is the exception, being affected only weakly. For $\ket{\mathbb{Z}_2}$ the figure also shows a chain of odd length, in which both end atoms are excited: removing the single facilitated edge atom of the even chain substantially deepens the minima for the bi-frequency drive, confirming that the bulk of $\ket{\mathbb{Z}_2}$ is protected at first order, while the residual infidelity stems from the facilitated channels opened by de-excitations in the bulk (section~\ref{sec:fpt:cascade}).

\section{Finite-size check for the density families}
\label{app:size}

\begin{figure*}
\centering
\includegraphics[width=\textwidth]{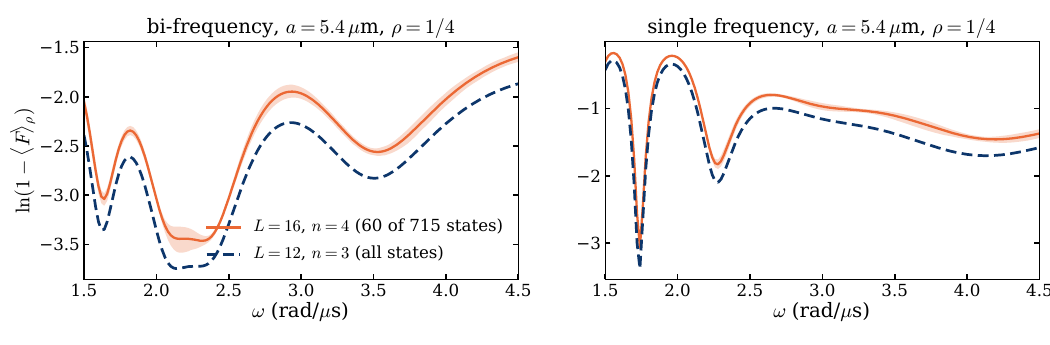}
\caption{\label{fig:sizecheck}Family-averaged infidelity at $\rho=1/4$ and $a=5.4\,\mu$m for bi-frequency (left) and single-frequency (right) driving. Dashed: all states with $n=3$ excitations for $L=12$. Solid: a random sample of $60$ out of the $715$ states with $n=4$ excitations for $L=16$; the shaded band is the $95\%$ bootstrap interval of the sample average. After subtraction of $\ln(16/12)$ the two curves agree to within the sampling uncertainty for bi-frequency driving and within $0.2$ for single-frequency driving.}
\end{figure*}

The density families of section~\ref{sec:families:n} are computed for $L=12$, the smallest length commensurate with all three densities. To test whether this length affects the structure of the family averages, we compare, for $\rho=1/4$ and $a=5.4\,\mu$m, the average over all states with $n=3$ excitations for $L=12$ with an average over a random sample of states with $n=4$ excitations for $L=16$ (figure~\ref{fig:sizecheck}). Since the number of states grows rapidly with $L$, the larger system is sampled rather than enumerated, and the sampling uncertainty is estimated by bootstrap resampling. For both protocols the two averages show the same minima at the same frequencies and differ by an almost uniform offset, with the larger system lying higher. This offset is close to $\ln(16/12)$, as expected from equation~\eqref{eq:F1}, according to which $1-F$ grows in proportion to the number of flippable atoms. Once it is removed, the remaining differences are within the sampling uncertainty for bi-frequency driving and small compared with the structure of the curves for single-frequency driving. The family averages at $L=12$ therefore capture the structure of the freezing landscape, while their overall level scales with system size.

\end{document}